\documentclass[preprint,preprintnumbers,amsmath,amssymb]{revtex4-1}

\usepackage{color}
\usepackage{titlesec}
\usepackage[]{graphicx}
\usepackage[resetlabels]{multibib}
\usepackage{latexsym}
\usepackage{natbib}
\usepackage{amsmath}
\usepackage[caption=false]{subfig}
\usepackage{multirow}
\usepackage{color}
\usepackage{xr}

\begin{document}

\title{Inferring Personal Economic Status from Social Network Location}

\author{Shaojun Luo$^{1}$, Flaviano Morone$^1$, Carlos Sarraute$^2$, Mat\'ias Travizano$^2$ and  Hern\'an A. Makse$^{1\dagger}$}

\affiliation{$^1$Levich Institute and Physics Department, City College of New York, New York, NY 10031 USA\\ $^2$Grandata Labs, 550 15th Street, San Francisco, CA 94103 USA\\ $^\dagger$ Corresponding author: hmakse@lev.ccny.cuny.edu}

\begin{abstract}
{\bf It is commonly believed that patterns of social ties affect individuals' economic status.  Here, we translate this concept into an operational definition at the network level, which allows us to infer the economic wellbeing of individuals through a measure of their location and influence in the social network. We analyze two large-scale sources: telecommunications and financial data of a whole country's population. Our results show that an individual's location, measured as the optimal collective influence to the structural integrity of the social network, is highly correlated with personal economic status. The observed social network patterns of influence mimics the patterns of economic inequality. For pragmatic use and validation, we carry out a marketing campaign that shows a three-fold increase in response rate by targeting individuals identified by our social network metrics as compared to random targeting. Our strategy can also be useful in maximizing the effects of large-scale economic stimulus policies.} \end{abstract}

\maketitle


\bigskip

\noindent {\bf Introduction} 

The long-standing problem of how the network of social contacts \cite{SciRef1,caldarelli,wasserman} influences the economic status of individuals has drawn large attention due to its importance in a diversity of socio-economic issues ranging from policy to marketing \cite{Granovetter,granovetter2005impact,Sci,pentland}. Theoretical analyses have pointed to the importance of the social network in economic life \cite{granovetter2005impact} as a medium to diffuse ideas \cite{smith2005networks,strang1998diffusion} through the effects of ``structural holes'' \cite{burt2009structural} and ``weak ties'' in the network \cite{Granovetter}. Likewise, research has recognized the positive economic effect of expanding an individual's contacts outside their own tightly connected social group \cite{SciRef1,SciRef2,SciRef6,zimmer1986entrepreneurship}. While previous work has established the importance of social network influence to economic status, the problem of how to quantify such correspondence via social network centralities or metrics \cite{wasserman,freeman} remains open. 

Studies employing mobile phone communication data and other social indicators have found a variety of network effects on socio-economic indicators such as job opportunities \cite{new1,new2}, social mobility \cite{new3,new4,new5}, economic development \cite{Sci,new6,new7,new8} and consumer behavior \cite{new9,new10}. Recent work also provides evidence of such effects on an individual's wealth, and highlights the need for better indicators \cite{new11}. Recently, a numerical study has tested the effect of network diversity on economic development \cite{Sci}. This study analyzed economic development defined at the community level. However, the question of how social network metrics may be used to infer financial status at the individual level\textemdash necessary, for instance, for micro-target marketing or social intervention campaigns\textemdash still remains unanswered. The difficulty arises, in part, due to the lack of empirical data combining an individual's financial information with the pattern of their social ties at the large-scale network level of the whole society. 

In this work, we address this problem directly by combining two massively-large datasets: a social network of the whole population of a Latin American country and financial banking data at the individual level. We discover that the optimal location of an individual in the network, which is measured by the collective influence metric (CI) \cite{CI}, is highly correlated with the individual's economic status at the population level: the larger the collective influence, the higher the socio-economic level.  The goodness of fit of this correlation can be as high as $R^2 =0.99$ when age is also included.  These results indicate that the optimal location in the social network measured by the collective influence metric can accurately predict socio-economic indicators at the personal level.

The top 1\% of the economic stratum has precise network patterns of ties formation showing relatively low local connectivity surrounded by a hierarchy of hubs strategically located in spheres of influence of increasing size in the network. Such a pattern is not observed in the rest of the population, in particular, in the bottom 10\% characterized by low values of collective influence.  Thus, the influence measured from social network patterns mimics the inequality observed in economic status \cite{stiglitz}.

We also find a high correlation between the link diversity of individuals and their financial status ($R^2 = 0.96$) employing the analysis based on network location and age. Analysis of the covariance suggests that the effect of network influence is significant and independent from other factors. We validate these results by carrying out a targeted marketing campaign in which we compare the response rate for different groups of people with different network locations. By targeting the group with the top collective influence values, the response rate can reach as high as 1\%; approximately three times the response rate found by random targeting and five times the response rate of the low collective influence people.

Thus, individuals with high socio-economic status (top 1\%) develop a very characteristic pattern of social ties as compared to the bottom 10\%. While this result may be expected, it is remarkable that the difference in pattern of social interactions between the rich and the poor can be precisely captured by a network metric measuring their collective influence in the social network \cite{CI}. The top socio-economic layer of society also represents the minimal set of people that provides integrity to the whole social network through their large collective influence. The fact that individuals of higher economic status are located in regions of large collective influence in the network elevates previous anecdotal evidence to a principle of network organization through the optimization of influence of affluent people affecting the structural integrity of the social network. At the same time, it suggests the emergence of the phenomenon of collective influence in society as the result of the local optimization of socio-economic interactions.

\bigskip

\noindent {\bf Results}

{\bf Network Construction}

The social network is constructed from mobile (calls and SMS metadata) and residential communications data collected for a period of 122 days (Supplementary Note \ref{sm1}, aggregated data at \url{kcorelab.com}). The database contains 1.10$\times$10$^8$ phone users. After filtering the non-human active nodes by a machine-learned model trained on human natural communication behavior (Supplementary Note \ref{sm2} , with Supplementary Figures \ref{fig:logiFit} -- \ref{fig:filResult}), we construct a final network of 1.07$\times$10$^8$ nodes in a giant connected component made of 2.46$\times$10$^8$ links. The ties, or links, in the network correspond to phone call communications, since we expect that communication patterns are indicative of an individual's location in the social network \cite{bar,martha,eagle2009inferring}. The financial cost of using phone services makes it possible that there is a systematic bias in how much wealthy individuals use the phone services relative to people that have less money to spend on phone calls. Although the effect might be limited (see Supplementary Note \ref{sm1}), we cannot rule out this possibility with the present data.

Financial status is obtained from the combined credit limit on credit cards assigned by banking institutions to each client. The credit limit is based on composite factors of income and credit history and therefore reflects the financial status of the individual (see discussion in Supplementary Note \ref{sm1}). The credit limit is pulled from an encrypted bank database and identified by the encrypted clients' phone numbers registered in the bank. Thus, we are able to precisely cross-correlate the financial information of an individual with their social location in the phone call network at the country level. There are 5.02$\times$10$^5$ bank clients who have been identified in the mobile network whose credit limit ranges from USD $\$50$ to \$3.5$\times$10$^5$ (converted from the country of study). Thus, the datasets are precisely connected providing an unprecedented opportunity to test the correlation between network location and financial status.

Despite the large scale of our data source, we note that working on a single specific country as in the present study is not enough to grant generality to our results. In order to test the general validity of the present results, access to other countries'whole-population-level communication and banking datasets would be needed. As more datasets become available, the generality of our results can be tested across different economic and social systems.

\begin{figure}
  \centering \includegraphics[width=\textwidth]{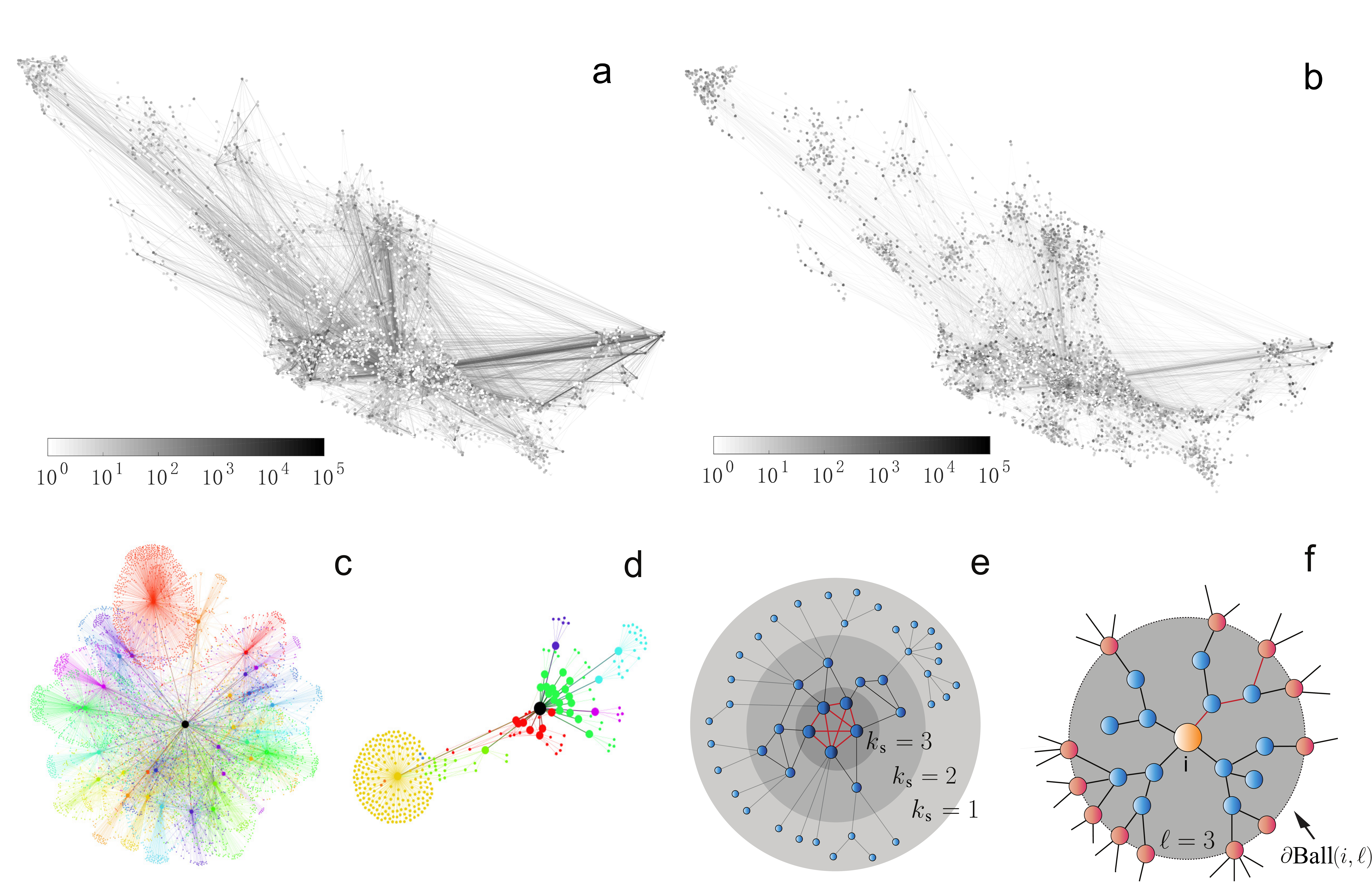}
   \caption{{\bf Patterns of network influence mimic patterns of income inequality.} Visualization of communication activity of the population in \textbf{(a)} the top 1\% (with credit limit larger than USD \$25,000, converted, in the country of study) and \textbf{(b)} bottom 10\% (with credit limit smaller than USD \$600, converted) of total credit limit classes. Links are between bank clients who have registered their zip code. Resolution of both plots is $1700\times1000$. The number of bank clients inside each community is reflected by the size of the node. Average credit limit is denoted by a node's grayscale. The color and thickness of the edges reflects the number of communication events between different communities. \textbf{(c)} Examples of the ego-network (extended to two layers) for an individual in the top 1\% wealthy class and \textbf{(d)} an individual in the bottom 10\% class.  The networks show two distinct patterns of social ties according to high and low economic status: the former is characterized by large CI, the latter by low CI. \textbf{(e)} Schematic representation of a network under k-shell decomposition \cite{kshell}. \textbf{(f)} Example of the calculation of CI. The collective influence Ball$(i, \ell)$ of radius $\ell=3$ around node $i$ is the set of nodes contained inside the sphere and $\partial {\rm Ball}$ is the set of nodes on the boundary (brown). CI is the degree-minus-one of the central node times the sum of the degree-minus-one of the nodes at the boundary of the sphere of influence.}
  \label{fig:demos}
  \end{figure}

Figures \ref{fig:demos}a and \ref{fig:demos}b show the communication patterns geolocalized across the country of individuals in the top 1\% and bottom 10\% of credit limits, respectively. The inequality in the patterns of communication between the top economic class and the lowest is striking and mimics the economic inequality at the country level \cite{stiglitz}. It is visually apparent that the top 1\% (accounting for 45.2\% of the total credit in the country) displays a completely different pattern of communication than the bottom 10\%; the former is characterized by more active and diverse links, especially connecting remote locations and communicating with other equally affluent people. Further results using entropy analysis also suggest that the network structure may be significantly different between the people in the top and bottom quantile rankings of credit limit (Supplementary Note \ref{smentropy} and Supplementary Table \ref{table:entropy}). Particular examples of the extended ego-networks for two individuals (with same number of ties) ranking in the top 1\% and bottom 10\% provide a zoomed in picture of such differences (Figs. \ref{fig:demos}c and \ref{fig:demos}d, respectively). The wealthiest 1-percenters have higher diversity in mobile contacts and are centrally located, surrounded by other highly connected people (network hubs). On the other hand, the poorest individuals have low contact diversity and are weakly connected to fewer hubs. The crux of the matter is to find a reliable social network metric to quantify this visual difference in the patterns of network structure between the rich and the poor, as we show next.

{\bf Network Influence and Financial Status}

Many metrics or centralities have been considered to characterize the influence or importance of nodes in a network \cite{wasserman,freeman,sen}. Here, we consider only those centralities that can be scaled up to the large network size considered here (Figs. \ref{fig:demos}e, \ref{fig:demos}f and Supplementary Note \ref{metrics}): (a) degree centrality $k_i$ (number of ties of individual $i$) is one of the simplest \cite{wasserman}, (b) PageRank, of Google fame \cite{pgrank}, is an eigenvector centrality that includes the importance of not only the degree, but also the nearest neighbors, (c) the k-shell index $k_\mathrm{s}$ of a node (Fig. \ref{fig:demos}e), i.e., the location of the shell obtained by iteratively pruning all nodes with degree $k \le k_\mathrm{s}$ \cite{kshell}, and (d) the collective influence of a node with degree $k_i$ (Fig. \ref{fig:demos}f) in a sphere of influence of size $\ell$ defined by the frontier of the influence ball $\partial {\rm Ball} (i,\ell)$, and predicted to be ${\rm CI}=(k_{i}-1)\sum _{j\in \partial {\rm Ball}(i,\ell)}{({k}_{j}-1)}$ by optimal percolation theory ~\cite{CI}. As opposed to the other heuristic centralities, CI is derived from the theory of maximization of influence in the network \cite{kempe}. The top CI nodes are thus identified as top influencers or superspreaders of information, and they do so by positioning themselves at strategic locations at the center of spheres surrounded by hubs hierarchically placed at distances $\ell$ (Fig. \ref{fig:demos}d). These collective influencers also constitute an optimal set that provides integrity to the social fabric: they are the smallest number of people that, upon leaving the network (a process mathematically known as optimal percolation \cite{CI}), would disintegrate the network into small disconnected pieces.

\begin{figure} 
  \includegraphics[width=\textwidth]{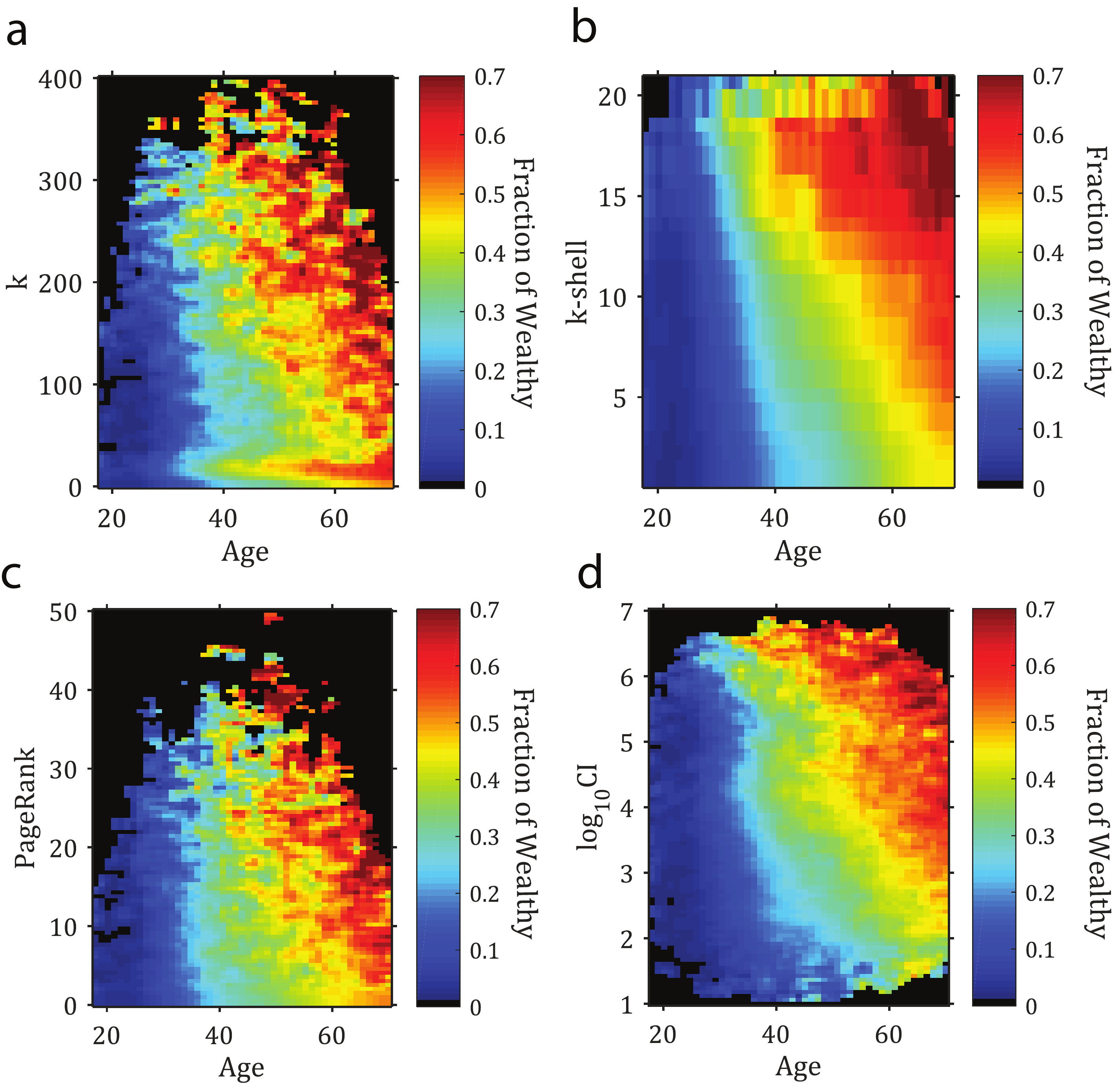}
  \caption{{\bf Fraction of wealthy individuals vs. age and network metrics.} Correlation between the fraction of wealthy individuals vs. age and \textbf{(a)} degree $k$ ($R^2=0.92$), \textbf{(b)} k-shell ($R^2=0.96$), \textbf{(c)} PageRank ($R^2=0.96$), and \textbf{(d)} $\log_{10}$CI ($R^2=0.93$). Only those groups with population larger than 20 are shown in the plot. The four metrics correlate well with financial status when considered with age. Further correlations are studied in Supplementary Note \ref{sm6}, indicating that CI could be considered as the most convenient metric out of the four due to its high resolution.}
  
  \label{fig:features}
  \end{figure}

By definition, all of the metrics have similarities (e.g., they are proportional to $k$, and PageRank and CI are based on the largest eigenvalues of the adjacency and non-backtracking matrices, respectively \cite{CI}), and indeed, we find that their values in the phone communications network are correlated (Supplementary Table \ref{table:tbl1}). More interestingly, Fig. \ref{fig:features} provides evidence of correlation of the four network metrics with financial status (ranked credit limit) when we control for age, indicating that the network location correlates with financial status. In this figure, we plot the fraction of wealthy individuals (defined as top 4th quantile, equivalent to a credit limit greater than USD \$4,000; see Supplementary Note \ref{sm4} for details about validation methods and \cite{eagle2009inferring}) in a sampling grid for a given value of age and social metric as indicated.

While all of the social metrics show correlations with financial status when considered with age (Fig. \ref{fig:features}), the question remains of which metric is the most efficient predictor. Strong correlations with economic wellness are observed for the feature pairs (age, k-shell) ($R^2 = 0.96$, Fig. \ref{fig:features}b) and (age, CI) ($R^2 = 0.93$, Fig. \ref{fig:features}d). Supplementary Note \ref{sm6} (Supplementary Figures \ref{fig:CLDist} -- \ref{fig:AVL}) provides further comparison when considering the metrics alone, indicating that k-shell and CI better capture the correlation with credit limit. Between these two metrics, CI guarantees a requirement for both strong correlation and sufficient resolution. K-shell cannot capture further details due to its limitation of values (k-shell ranges from 1 to 23, dividing the whole population into this small number of shells with a typical shell containing tens of millions of people), while CI spans over seven orders of magnitude; see Supplementary Figure \ref{fig:Distri}. This high resolution implies that CI is a more accurate social signature for the financial status of the individuals. According to its definition (Fig. \ref{fig:demos}d), a top CI node is a moderate to strong hub surrounded by other hubs hierarchically placed at distance $\ell$. However, we emphasize that CI is just a useful strategy for the reasons shown above, and by no means the only or best strategy to correlate the wealth of individuals and their network influence.

While the theory behind CI is a global maximization of influence, CI represents the local approximation to this global optimization. Thus, CI represents a balance between a global optimization and its local approximation, taking into account the first 2 or 3 layers of neighbors via the parameter $\ell$, which represents the size of the sphere of influence used to define the importance of a node, Fig. \ref{fig:demos}d. By changing $\ell$, we discover that CI with $\ell = 2$ is sufficient to capture the correlation between network influence and wealth (Supplementary Figure \ref{fig:CI123}).

\begin{figure} 
  \centering
  \includegraphics[width=\textwidth]{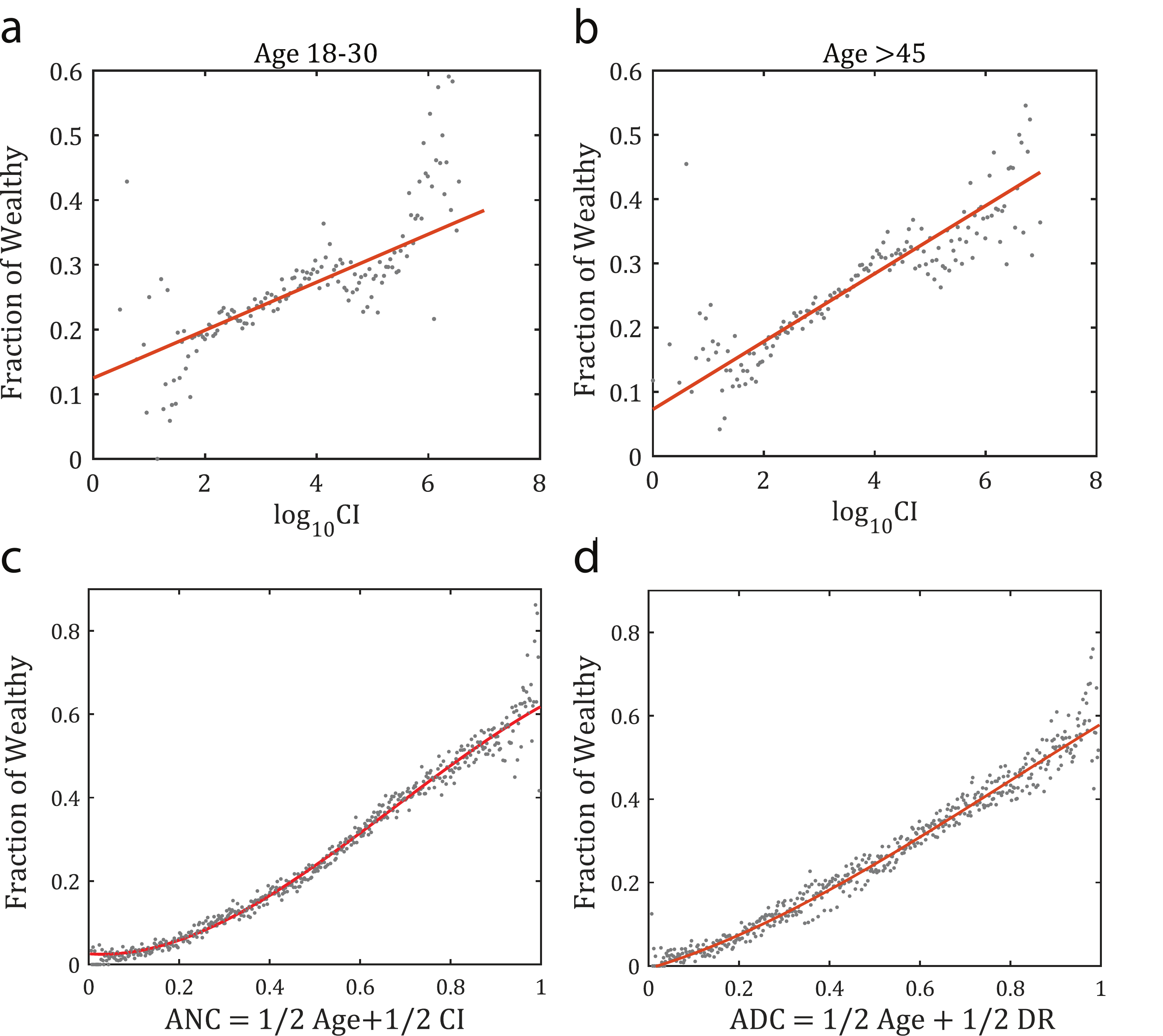}
  \caption{{\bf Fraction of wealthy individuals over
      different age and composite ranking groups.} Correlation between
    the fraction of wealthy individuals as given by the top 25\%
    credit limit and CI in different age groups of \textbf{(a)} 18-30,
    \textbf{(b)} $>$45. Correlations between top economic status and
    large collective influence as determined by CI values in different
    ages are significant in all age groups, while the slope of the
    linear regression is larger in the older group (0.053 compared to
    0.037). \textbf{(c)} Age-network composite ranking ANC = 1/2 Age
    + 1/2 CI, and \textbf{(d)} age-diversity composite ranking ADC =
    1/2 Age+ 1/2 DR. By combining the network metrics with age into a
    composite index, the chance to identify people of high financial
    status reaches $\sim 70\%$ for high values of the composite. Both
    $R^2$ show a high level of correlation ($R^2$ = 0.99 and 0.96 for
    ANC and ADC, respectively), making both composites good predictors
    of wealth in practical applications.}
  
  \label{fig:ageCellCIinout}
  \end{figure}

To track the effect of CI independently of age we investigate the effects of CI inside two specific age groups in Figs. \ref{fig:ageCellCIinout}a and \ref{fig:ageCellCIinout}b. In both age groups, high CI is always accompanied by a higher population of wealthy people. A relatively smaller slope in age group $<$30 suggests that the CI network effect is more sensitive for older people with more mature and stable economic levels, than for younger people (see in Supplementary Figure \ref{fig:slopes}). When we combine age and CI quantile ranking into an age-network composite: ANC = $\alpha$ \, Age + $(1-\alpha)$ CI, with $\alpha=0.5$, a remarkable correlation ($R^2 = 0.99$, Fig. \ref{fig:ageCellCIinout}c) is achieved. By combining network information with age, the probability to identify individuals with a high credit limit reaches $\sim70\%$ at the highest earner level. Such a level of accuracy renders the model practical to infer individuals' financial fitness using network collective influence as we show next.

{\bf Validation by Marketing Campaign}

\begin{figure} 
  \includegraphics[width=\textwidth]{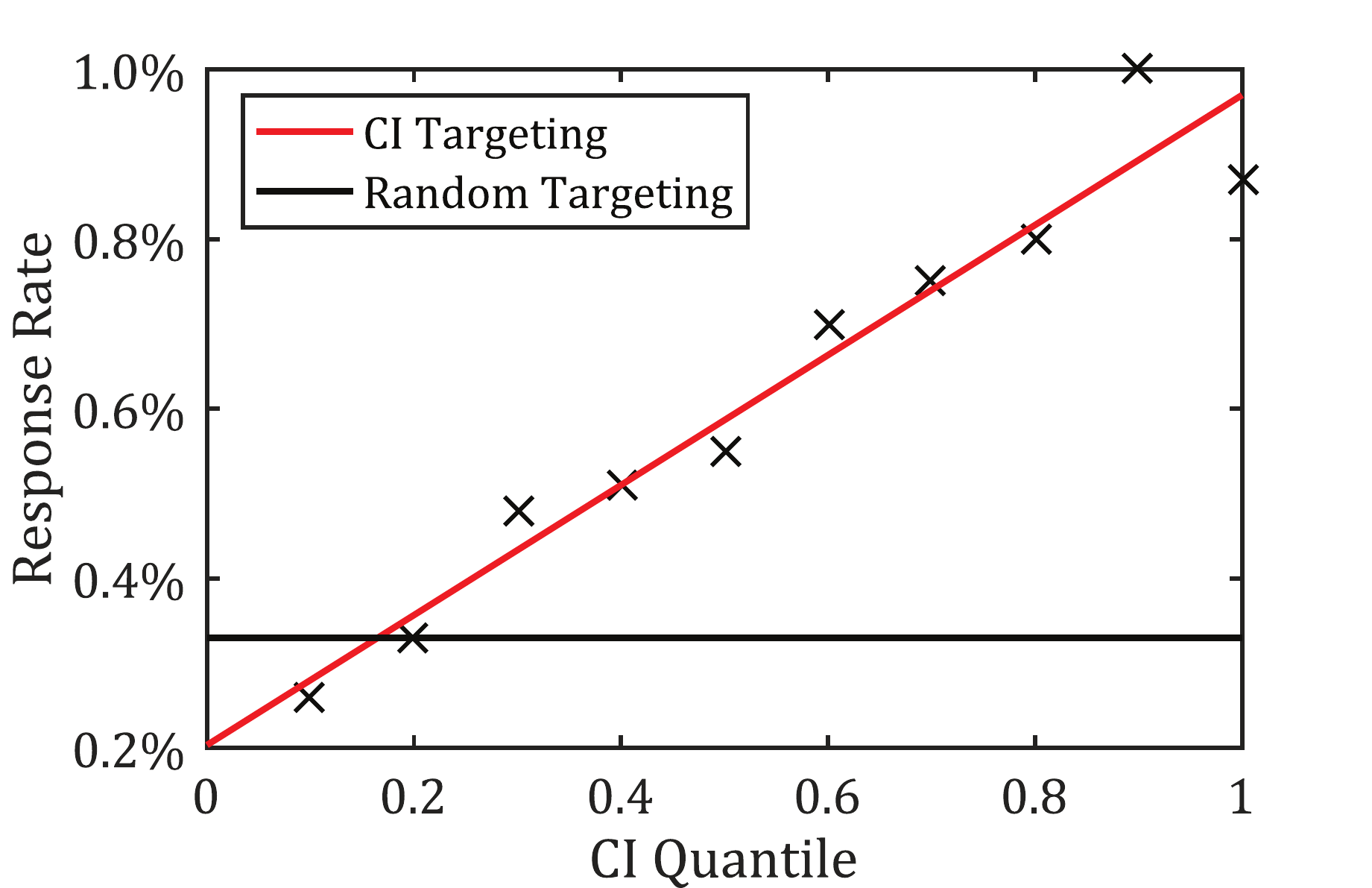}
  \caption{{\bf Response rate vs. CI quantile in the
      real-life CI-targeted marketing campaign.} The response rate
    increases approximately linearly with CI ranking. The CI-targeted
    campaign shows a three-fold gain for the top influencers with high
    CI, as compared with a campaign targeting a randomized control
    group.}
  
  \label{fig:response}
  \end{figure}

To validate our strategy we perform a social marketing campaign whose
objective is the acquisition of new credit card clients, by sending
messages to affluent individuals (as identified by their CI values)
and inviting the recipients to initiate a product request (see
Supplementary Note \ref{smmarket}). We note that in this experiment
we use an independent dataset from a different time frame, and we use
only the CI values extracted from the network to classify the targeted
people. Specifically, we use the communications network resulting from
the aggregation of calls and SMS exchanged between users over a period
of 91 days. The resulting social network contains 7.19 $\times 10^7$
people and 3.51 $\times 10^8$ links. The campaign was conducted on a
total of 656,944 people who were targeted by an SMS message offering
the product according to their CI values in the social network. We
also sent messages to a control group of 48,000 people, chosen
randomly. To evaluate the campaign, we measured the response rate,
i.e., the number of recipients who requested the product divided by
the number of targeted people, as a function of CI. In the control
group, the response rate to the messages was 0.331\%. Our results show
that groups of increasing CI show an increase in their response rate,
with a sound three-fold gain in the rate of response of the top
influencers (as identified by top CI values) when compared to the
random case.  When we compared the response of the high CI to the
lowest CI people, the response rate increased five-fold.  The results
of the experiment are summarized in Table \ref{tab:markcampaign} and
in Fig. \ref{fig:response}.

{\bf Analysis of Covariance}

We note that our validation is indirect since it is not a direct prediction of financial status, but a rate of successful response to a marketing campaign. This success rate may actually depend on a number of other factors that may correlate with the network centrality. Thus, the CI metric may not necessarily be the only cause of the success rate of the targeted campaign (for instance, geographical location may be also important). To address this point, we perform an Analysis of Covariance (ANCOVA) \cite{ANCOVA} on all of the features to which we have access (age, gender and registered zip code) to test the variance caused by the network metrics and other factors (details in Supplementary Note \ref{sm4} and Supplementary table \ref{table:corr}). ANCOVA shows that the effects of the network metrics are independent from those of the other factors. The correlation between the CI and the fraction of wealthy people is positive and significant ($p<0.001$) in all groups of geographical communities, across genders, and among all age groups older than 24 years (Supplementary Figure \ref{fig:slopes}). The same significant results are also obtained under different thresholds of wealth. Such significant and robust network effects imply that network metrics may be a potential indicator for financial status.

{\bf Network Diversity and Financial Status}

Our combined datasets also offer the possibility to test the importance of the diversity of links, as measured by ties to distant communities in the network not directly connected to an individual's own community, at the level of single individuals \cite{Granovetter,granovetter2005impact,Sci}. To this end, we first detect the communities in the social network by applying fast fold modularity detection algorithms (Supplementary Note \ref{sm7} and Supplementary Figure \ref{fig:distComW2}) \cite{comDtct,comDtct12}. The diversity of an individual's links can be quantified through the diversity ratio DR $=W_{\rm out}/W_{\rm in}$ \cite{burt2009structural}, defined as the ratio of total communication events with people outside their own community, $W_{\rm out}$, to those inside their own community, $W_{\rm in}$. This ratio is weakly correlated to CI ($R =0.4$), suggesting that it captures a different feature of network influence. We implement the same statistics of composite ranking as before, resulting in an age-diversity-composite ADC = $\alpha$ \, Age + $(1-\alpha)$ DR, with weight $\alpha=0.5$. The result (Fig. \ref{fig:ageCellCIinout}d) shows that ADC correlates with individual financial wellbeing, generalizing the aggregated results in \cite{Sci} to the individual level. Thus, the social metrics considered, DR and CI, express the fact that higher economic levels are correlated with the abilities to communicate with individuals outside one's local tightly-knit social community, a measure of Granovetter's ``strength of weak ties'' principle \cite{Granovetter} and to position oneself at particular network locations of high CI that are optimal for information spreading and structural stability of the social network. We note that no causal inference can be established with the present data.

\bigskip

\noindent {\bf Discussion}

This result highlights the possibility of predicting both financial status and benefits of socially-targeted policies based on network metrics, leading to tangible improvements in social marketing campaigns. The high performance of CI among network metrics also suggests the possible role of accessing and mediating information in financial opportunity and wellbeing \cite{granovetter2005impact}. This has an immediate impact in designing optimal marketing campaigns by identifying the affluent targets based on their influential position in a social network. This finding may be also raised to the level of a principle, which would explain the emergence of the phenomenon of collective influence itself as the result of the local optimization of socio-economic interactions.

\bigskip

\noindent {\bf Methods}

{\bf Code Availability}

Source code of the Collective Influence algorithm is available at the website \url{http://www-levich.engr.ccny.cuny.edu/~jmao/CI/CI_HEAP.c}. Other source code is available upon request to the authors. 

{\bf Data Availability}

The datasets generated during and/or analyzed during the current study are not publicly available for privacy reasons, but are available from the corresponding author on reasonable request.

\newpage

\bibliography{shaojun26_arXiv}
\bibliographystyle{naturemag}

\bigskip

{\bf Acknowledgments} 

This work was funded by ARL Cooperative Agreement Number
W911NF-09-2-0053 (the ARL Network Science CTA), NIH-NIGMS
1R21GM107641, and NSF-PoLS PHY-1305476. We thank B. Min for
discussions.

{\bf Author Contributions}

H. A. Makse directed the project. S. Luo developed the data and
network analysis. F. Morone performed the data validation and provided
the support on network theory. C. Sarraute and M. Travizano collected
and processed all original data. All authors contributed in writing
the paper.

{\bf Competing Interests}

The authors declare no competing financial interests.
\clearpage

\clearpage

\begin{table}
  \caption{{\bf Results of the real-life marketing campaign.}
    Individuals (``Count'') were targeted according to their quantile
    CI ranking in the whole social network obtained from phone
    communications activity. The response to the campaign (``Answered
    Yes'') was computed to calculate the Response Rate.  }
  \begin{tabular}{| c | c | c | c | c| }
    \hline
   CI range & Count & Quantile & Answered Yes & Response Rate\\
    \hline \hline
   [0,48] & 66495 & 0.1 & 170 & 0.26\%\\
    \hline
   (48, 246] & 65164 & 0.2 & 218 & 0.33\%\\
    \hline
   (246, 600] & 65961 & 0.3 & 316 & 0.48\%\\
    \hline
   (600, 1144] & 65376 & 0.4 & 332 & 0.51\%\\
    \hline
    (1144, 1992] & 65477 & 0.5 & 363 & 0.55\%\\
    \hline
    (1992, 3408] & 65477 & 0.6 & 458 & 0.70\%\\
    \hline
    (3408, 6032] & 65736 & 0.7 & 493 & 0.75\%\\
    \hline
    (6032, 11772] & 65641 & 0.8 & 555 & 0.8\%\\
    \hline
    (11772, 28740] & 65683 & 0.9 & 657 & 1.0\%\\
    \hline
    (28740, 2719354] & 65683 & 1.0 & 573 & 0.87\%\\
    \hline
  \end{tabular}

\label{tab:markcampaign}
\end{table}

\clearpage

\captionsetup[table]{name=Supplementary Table}
\captionsetup[figure]{name=Supplementary Figure}

\setcounter{figure}{0}
\setcounter{table}{0} 
\setcounter{section}{0}     

\titleformat{\section}{\bf}{}{0em}{Supplementary Note \arabic{section} - }
\renewcommand{\theequation}{\arabic{equation}}
\renewcommand{\thefigure}{\arabic{figure}}
\renewcommand{\thetable}{\arabic{table}}
\renewcommand{\thesection}{\arabic{section}}
\renewcommand{\figurename}{Supplementary Figure}
\renewcommand{\tablename}{Supplementary Table}

\bibliographystyle{naturemag}

\newenvironment{blockquote}{%
  \par%
  \medskip
  \leftskip=2em\rightskip=2em%
  \noindent\ignorespaces}{%
  \par\medskip}

\centerline{\bf SUPPLEMENTARY INFORMATION}
\bigskip

\section{Datasets}
\label{sm1} 

The present framework and data acquisition have gone through an extensive process of revision and approval that took more than one year and have IRB approval Protocol No. 2016-1418 at City University of New York.

In the framework of the study, the private and/or sensitive information of the telecommunications company clients was protected. In particular, the Bank didn't gain access to any individual information about the telecommunications company users. Similarly, the private information of the Bank's clients was protected in the framework of the study. In particular, the telecommunications company didn't have access to the individual information of the Bank's clients. The variables shared were revised to guarantee that the privacy of clients was protected.

All of our datasets are encrypted and securely stored. The mobile dataset consists of records of phone calls and SMS (short message service) metadata which was collected from clients of a major operator of a Latin American country. The dataset is anonymized. All the data are encrypted and stored in a server secured by enterprise-grade firewall. The records cover a period of 122 consecutive days. Each phone number was encrypted by a high level of hashing in order to eliminate all possible access to personal information. For our purposes, each CDR (Call Detail Record) is represented as a tuple $\langle$x, y, t, dur, d, l$\rangle$, where x and y are the encrypted phone numbers of the caller and the callee, t is the date and time of the call, dur is the duration of the call, d is the direction of the call (incoming or outgoing, with respect to the mobile operator client), and l is the location of the tower that routed the communication. Similarly, each SMS metadata record is represented as a tuple $\langle$x, y, t, d, l$\rangle$. We constructed a social network $G=(N,E)$ based on the phone call and SMS traffic. Both reciprocal and non-reciprocal links are preserved for further processing.

In inferring the real social network from the mobile network, we take the assumption that the communication demands are rigid against the cost, which is usually affordable to most families ($\sim$USD \$17 monthly cell phone service fee vs. $\sim$USD \$600 monthly income in the year data was collected, respectively). Thus, the direct impact of an individual's financial status on the communication structure evidenced in the mobile phone network might be limited. However, the financial cost of using phone services makes it possible that there is a systematic bias in how much wealthy individuals use the phone services relative to people that have less money to spend on phone calls. At this point, with the present data, we cannot rule out this possibility.

The financial dataset from a major bank in the same country was collected during the same time period as the mobile dataset. These data record financial details of $1.23 \times 10^6$ clients assigned unique anonymized identifiers over the same three-month period as the mobile network. The dataset consists of records of the bank clients' age, gender, credit score, total transaction amount during each billing period, credit limit of each credit card, balance of cards (including debit and credit), zip code of billing address, and encrypted registered phone number. A subset of $5.02\times 10^5$ clients have an encrypted mobile phone number, thus enabling them to be matched with the mobile communication dataset. The phone numbers are encrypted in the same way as in the mobile dataset, which guarantees that the two datasets are matched. Excluding the information on credit lines, all other personal information is erased. We sum up the credit limits of all the credit cards of each account owner to represent the total credit limit of each individual.

In the absence of direct access to an individual's income and total assets, evaluating an individual's financial status remains an open question. In this dataset, we can access the following factors:

Transaction amount, which also directly reflects the individuals' consumption patterns. However, since it is common that one holds multiple accounts in different banks, and some of these may not be used at all, records in only one bank might not correctly reflect the real spending ability of an individual. Similar reasoning can be applied to total credit card balance per month, which could also lose its ability to measure one's financial status.

Credit scores assigned to individuals by credit scoring agencies are
also good indicators of financial status. However, the values of
credit scores are quite limited, ranging from 300 to 850. This limited
range makes the credit score a low-resolution indicator of wealth that
does not allow us to correctly classify a large number of people into
well-defined financial classes. On the other hand, the credit limit
ranges over three orders of magnitude, allowing us to correctly
classify the entire population. Considering the weaknesses of the
other features, total credit limit is the most convenient measure of
personal financial status in the present dataset.

Instead of transaction amounts and credit scores, we choose the total credit limit which is assigned by the bank after comprehensive evaluation of an individual's financial status, as a proxy for financial status. Since detailed information on how the credit limit is assigned is not provided, there are several possible factors that could cause bias in inferring an individual's real economic status. These include the delay of credit limit in reflecting a change in an individual's financial status, possible correlation with the age of the account, and so on. In fact, the credit limit might be capturing the amount of information the bank has about the customer, instead of his/her actual income.

\section{Removing non-human-operated lines}
\label{sm2}

Inferring social network structure through mobile phone data requires the removal of lines operated by non-humans. Due to privacy restrictions, we could not filter business landlines and spawn spreaders at the outset. Several ways of filtering the landlines were applied in previous works, including setting a cut-off threshold degree \cite{Sci} or only considering reciprocal phone calls \cite{bar}. However, these methods usually also cut off some important human communication behavior in that particular window of observation. All communication events should be considered in evaluating the social network. Therefore, the key problem is to find a method to distinguish human- and non-human-operated lines while retaining maximal information about individuals' communication patterns.

Although we do not have the human/non-human label for the totality of the phone lines, which could separate at the outset the non-human-operated lines, we are in possession of the set of phone numbers registered with the bank dataset. These human-operated lines provide the possibility of supervising a machine learning process to learn the human behavior that separates them from robots and non-human-operated lines. We set up a hypothesis test by modeling the human-operated lines based on several variables. We first cluster the human-operated lines in a hyperspace. A new unlabeled node will be assigned a p-value according to its distance to the cluster. By carefully choosing a threshold of the p-values, we can label the node according to whether we accept or reject the hypothesis that the line is operated for personal use.

A training set consisting of the phone lines in the bank database ($1.23 \times 10^6$ nodes), which is around 1\% of all of the data in the entire network ($1.10 \times 10^8$ nodes), was set up. We define a call or message from phone number $i$ to $j$ as a `communication event,' and denote the total number of communication events on the link as $W_{i\rightarrow j}$. The key assumptions of the model are the
following:

1. Communication between lines of personal use is usually (but not always) reciprocal. This means that the fraction of paired communication events on human-operated lines is generally higher than that of unpaired ones. Namely, it suggests that although communication load difference $D_i$ on every line:
  \begin{equation}
    D_i=\left|\sum _{j\in \partial i} W_{i\rightarrow j} - \sum _{j\in
      \partial i} W_{j\rightarrow i}\right|
  \end{equation}
should increase with degree $k$, it should be bound by an upper limit in the case of human-operated lines. Numbers operated for non-personal use like business hubs and spawn spreaders may have very large $D_i$ because they are usually operated only for sending or receiving phone calls independently, but not for both at the same time.

2. Other types of business hubs may have large numbers of paired communications despite their limited $D_i$. These business hubs include the phone numbers for company landlines, roadside assistance, or other services requiring instant follow-up by the recipient of the phone call. To filter out these hubs we assume that the paired communication:
  \begin{equation}
    R_i=\sum _{j\in \partial i}\min(W_{i\rightarrow j},W_{j\rightarrow
      i})
  \end{equation}
also increases with $k$, but is limited for lines for personal use. The decay of the tail is supposed to follow a power-law due to the preferential attachment rule \cite{bar}. 

The last assumption is:
3. Most phone numbers in the network are for personal use, which results in the number of non-human-operated lines being small.

After we introduce these basic assumptions, empirical analysis can be applied to build a model describing human-operated line behavior. The model simplifies to a parametric probability distribution depending on two random variables $D_i$ and $R_i$, and a variable maximum degree $k$ which controls the parameters. Under the preferential attachment rule of assumption 2, it is reasonable to assume the distributions of both $D_i$ and $R_i$ for a given $k$ deviate from a maximum entropy distribution and show a power-law tail. A good approximation is the log-logistic distribution:
  \begin{equation}
    P(D_i|k)\sim LL(d_i,\alpha_{\mathrm D}(k),\beta_{\mathrm D}(k)), 
  \end{equation}
and
  \begin{equation}
P(R_i|k)\sim LL(r_i,\alpha_{\mathrm R}(k),\beta_{\mathrm R}(k)),
  \end{equation}
where
\begin{equation}
LL(x,\alpha(k),\beta(k))=\frac{(\beta/\alpha)(x/\alpha)^{\beta-1}}{[1+(x/\alpha)^{\beta}]^2}.
\end{equation}
 
\begin{figure} 
\includegraphics[width=\textwidth]{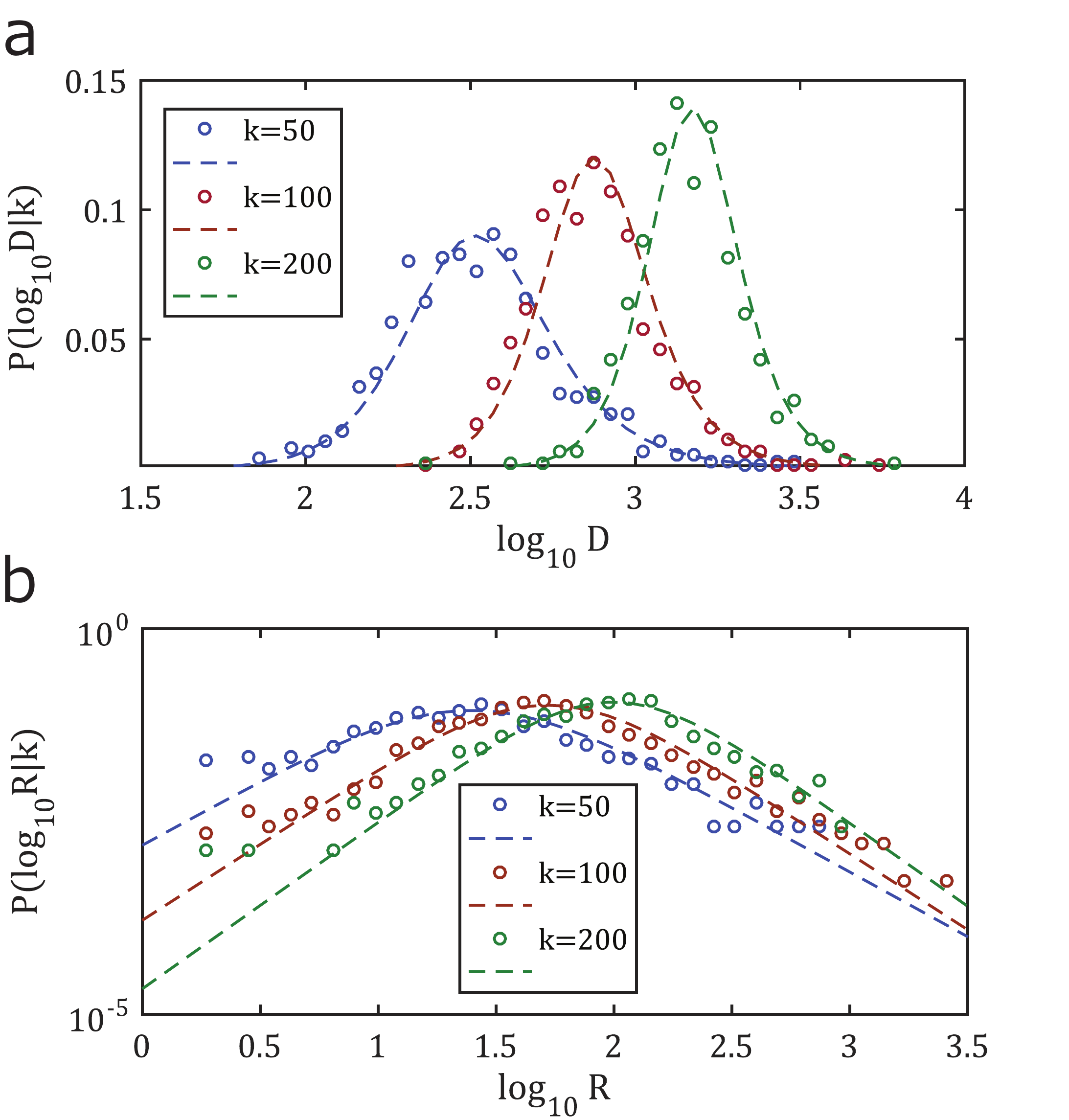}
\caption{{\bf Logistic fitting result for $k$ = 50, 100 and 200.} The result of paired communication $R$ is presented in log-log scale in order to highlight the fitting for the exponential tails.}
  
\label{fig:logiFit}
\end{figure}

\begin{figure} 
\includegraphics[width=\textwidth]{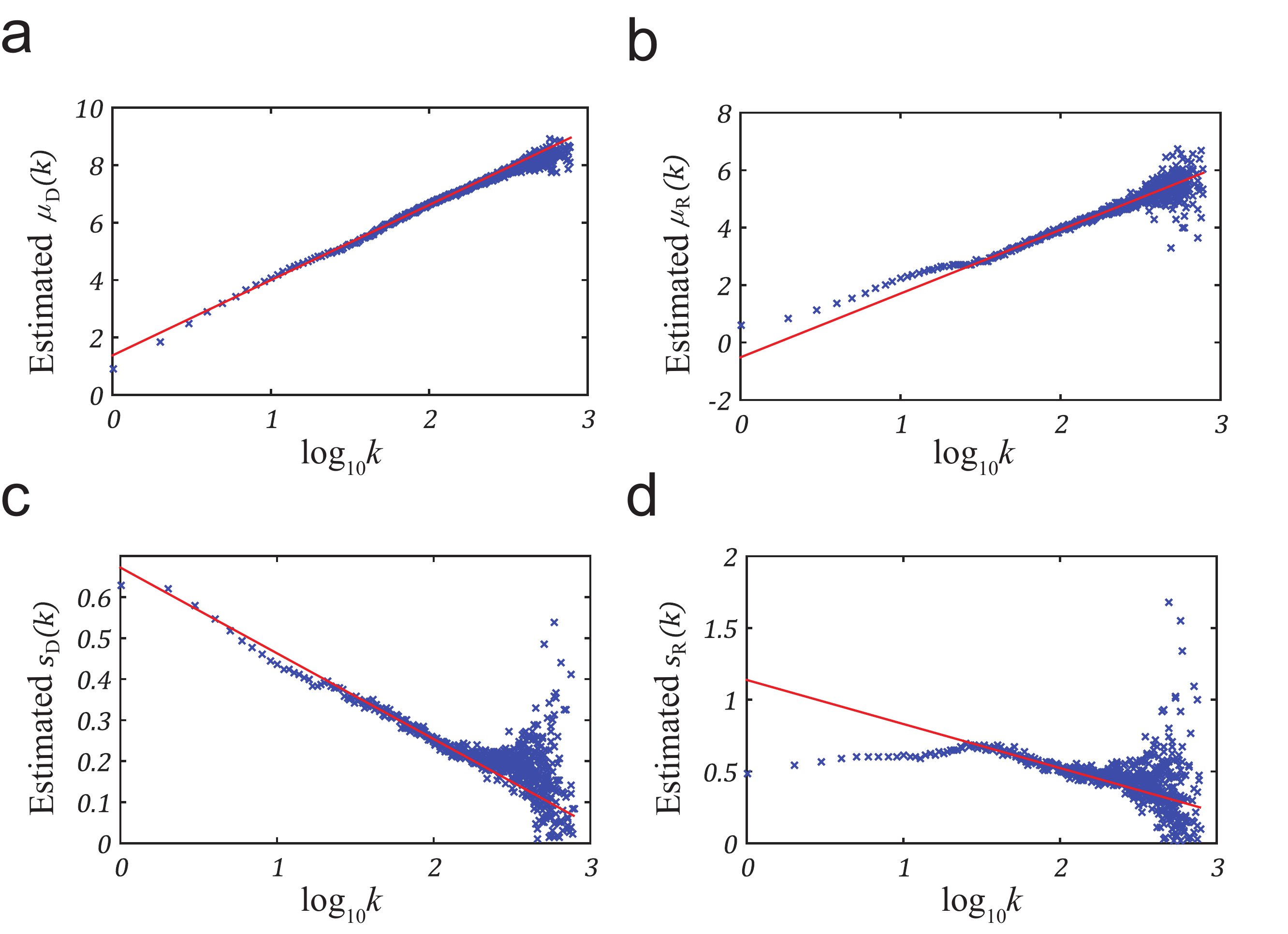}
\caption{{\bf Scaled parameter estimation and its linear fitting:} {\textbf (a)} $\hat{\mu}_\mathrm{D}(k)$, {\textbf (b) } $\hat{s}_\mathrm{D}(k)$, {\textbf (c) } $\hat{\mu}_\mathrm{R}(k)$, {\textbf (d) } $\hat{s}_\mathrm{R}(k)$.}

\label{fig:fitPar}
\end{figure}

This also suggests the logarithm of both metrics follows a normal-like but exponential tailed logistic distribution:
\begin{equation}
P(\log D_i|k)\sim L(d_i,\mu_{\mathrm D}(k),s_{\mathrm D}(k)), 
\end{equation}
and
    \begin{equation}
P(\log R_i|k)\sim L(r_i,\mu_{\mathrm R}(k),s_{\mathrm R}(k)),
  \end{equation}
where 
  \begin{equation}
L(x,\mu(k),s(k)) = \frac{1}{4s(k)} \mathrm{sech}^2\left(\frac{x-\mu(k)}{2s(k)}\right),
  \end{equation}

with $\mu(k)=\log(\alpha(k))$, and $s(k)=\frac{1}{\beta(k)}$. Based on the knowledge we have, this distribution is the best choice even though we cannot precisely provide an exact fitting. However, the fitting results strongly support the approximation geometrically (Supplementary Figure \ref{fig:logiFit}). The model involves four parameter sequences: $\hat{\mu}_\mathrm{D}(k)$, $\hat{s}_\mathrm{D}(k)$ and $\hat{\mu}_\mathrm{R}(k)$, $\hat{s}_\mathrm{R}(k)$. To determine the function of dependency, we pick the interval $k$ = 40 to 160. We consider this a normal range of degrees wherein the nodes are almost all human-operated to fit the trend of $\mu$ and $s$. Adequate numbers of observers in each degree division guarantee the reliability of the results. The estimated $\hat{\mu}_\mathrm{D}(k)$, $\hat{s}_\mathrm{D}(k)$ and $\hat{\mu}_\mathrm{R}(k)$, $\hat{s}_\mathrm{R}(k)$ can be simply described by linear models within this range (Supplementary Figure \ref{fig:fitPar}, $R^2>0.98$). The relations are then used to predict parameters under other degree ranges.

After validating the assumptions, we are able to implement the learning process by performing a hypothesis test:

1. Fit the model of training data and get the sequence of estimated $\hat{\mu}_\mathrm{D}(k)$, $\hat{s}_\mathrm{D}(k)$, $\hat{\mu}_\mathrm{R}(k)$, and $\hat{s}_\mathrm{R}(k)$.

2. For each node $i$ with given difference $d_i$, number of communication pairs $r_i$ and degree $k_i$, calculate the p-value of $p_\mathrm{D}(i)=P(D<d_i |k_i)$, and $p_\mathrm{R}(i)=P(R<r_i|k_i)$.

3. Set a threshold $p$ using the following test to classify the nodes:

If:
  \begin{equation}
    p<p_\mathrm{D}(i)<1-p \quad \land \quad p<p_\mathrm{R}(i)<1-p
  \end{equation}
then $i$ is a human-operated line. Otherwise a p-value outside the range defined above will be rejected by the null hypothesis: $H_0 \rightarrow i$ is a human-operated line. It will be labeled as a non-human-operated business hub due to its extraordinarily unbalanced communication pattern or large volume of communication events.

\begin{figure} 
\includegraphics[width=\textwidth]{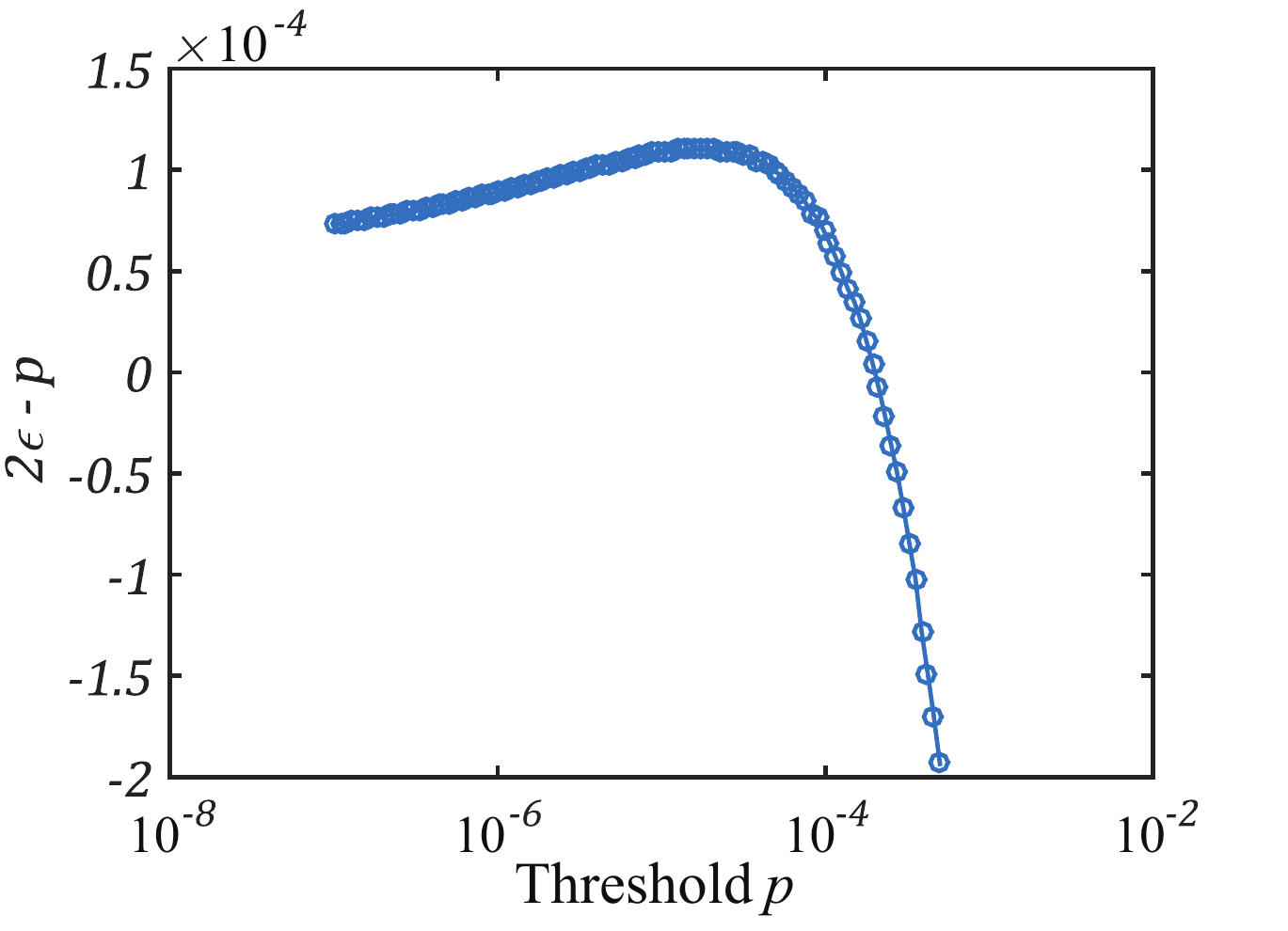}
\caption{{\bf Number of outliers $\epsilon-2p$ vs cut-off threshold $p$.} Maximum is reached when $p \sim 1.6 \times 10^{-5}$.}

\label{fig:optp}
\end{figure}

Last but not least, the threshold $p$ should be optimized. Suppose the network follows the exact distribution given by the model above. The fraction of outliers (non-human-operated lines) $\epsilon$ is exactly $2p$. The difference $\epsilon-2p$ can be approximately regarded as the number of non-human-operated lines or `outliers'. Supplementary Figure \ref{fig:optp} is the plot of $p$ over $\epsilon-2p$. A maximum is reached when $p \sim 1.6 \times 10^{-5}$. At that point, the filter is the most sensitive to detecting outliers since it covers the boundary of human- and non-human-operated nodes.

The result of data filtering is shown in Supplementary Figure \ref{fig:filResult}. The final network has $1.07\times10^8$ nodes (97.27\% of the total data) and $2.46\times10^8$ links. There are $4.51\times10^7$ reciprocal social ties. The size of the giant connected component is 99.2\% and the average degree is 4.7. The maximum degree $k$ is 1056 and the maximum total communication load of a single node is $\sim 10K$ including messages and calls, which is reasonable for a person who is active in business contacts during a three-month period.

\section{Entropy Analysis}
\label{smentropy}

In order to explore the structural differences between people with different levels of credit limits, we performed an entropy analysis. First, we choose people within the top $5\%$ and bottom $5$ to $10\%$ credit limit percentiles, representative of the wealthy and poor populations respectively. Then, we randomly divided both groups into 20 small subgroups where each subgroup contained $N(0)\sim2700$ bank clients. Next, we expanded each subgroup's contacts by a distance $\ell$ to get a subnetwork and clustered the nodes in the subnetwork through modularity analysis (Supplementary Note \ref{sm6}) into different communities, finally counting the number of nodes inside each community ($n_i$). The entropy of this subnetwork is defined as:

\begin{equation}
S=-\sum_i p_i \log p_i,
\end{equation}
where $p_{i} = \frac{n_i}{\sum_i n_i}$ is the fractional size of community $i$. Also, we introduced two indicators: (1) $R_\mathrm{n}(\ell) = N(\ell)/N(0)$, which is the ratio between the size of the augmented network $N(\ell)$ and the size of the initial subgroup $N(0)$, and (2) $R_\mathrm{c}(\ell) = C(\ell)/C(0)$, where $C(\ell)$ is the number of communities in the augmented network and $C(0)$ is the number of communities in the initial subgroup. Supplementary Table \ref{table:entropy} shows the results of entropy $S$, $R_\mathrm{n}(\ell)$ and $R_\mathrm{c}(\ell)$ across an average of 20 subgroups, with uncertainties.

The entropy in subnetworks generated from the poor population is higher than in subnetworks generated from the wealthy population, while the numbers of both the total communities and nodes are smaller. This suggests that the sizes of the communities in the subnetwork of poor people are relatively more balanced than in the wealthy population. Namely, wealthy people are more likely to form larger and more closely-connected communities which result in relatively low entropy. The result of $R_\mathrm{n}$ and $R_\mathrm{c}$ shows the significant difference between the size and diversity of the subnetworks of the wealthy and poor populations. By expanding their contacts, people with higher credit limits `collect' more people and more communities. Such differences exist even when we increase the value of $\ell$ to 4. The result of the entropy analysis implies that the network structure of these two groups may be significantly different. Wealthy people have higher diversity in mobile contacts and are centrally located, surrounded by other highly-connected people (network hubs).

Entropy analysis results also provide evidence of homophily, which implies that there exists a higher probability that two wealthy individuals are connected than that a wealthy individual and an extremely poor individual are connected. Since society is known to have this strong stratification property embedded in social networks, we would expect that this feature is expressed in our network. For example, if wealth implies higher degree, then homophily will lead to degree correlations, higher k-shell scores for wealthy individuals, and higher CI. Thus, part of the effect we observe in the present study might be due to the effects of homophily. However, the exact picture of how homophily affects the wealthy population is still to be discovered.

\section{Social Network Metrics}
\label{metrics}

In order to capture the analytical evidence describing the effects shown in Figs. \ref{fig:demos}a--d, we introduce four different metrics to evaluate network influence \cite{wasserman,freeman}.

1. Degree centrality $k_i$ is the simplest evaluation of an individual's local contact size. It requires minimum information and is easy to calculate. Other centralities such as betweenness centrality cannot be efficiently calculated in our networks due to their nonlinear running times with system size.

2. k-core and k-shell index $k_\mathrm{s}$ \cite{kshell} capture the centrality of a node in the global network by the method of k-shell decomposition. In this method, nodes are removed iteratively if their degree $k_i < k$ until all the remaining nodes have degree equal to or greater than $k$. These nodes remain in the k-core of index $k$. The largest k-core a node can hold is the k-shell index $k_\mathrm{s}$, which means the node is in the `shell' of the $k$'th core but outside the $k+1$'th core. The k-shell or k-core number is a global metric. It has been proven efficient in identifying single influencers through the SIR model \cite{kshell}. The k-shell index requires the overall information of the network. It is a quantity that does not allow one to classify the nodes with high resolution: there usually exist a few k-shells in the whole system, each containing many of the nodes in the network. Fig. \ref{fig:demos}c is a schematic example of a k-shell in a network.

3. PageRank \cite{pgrank} is as eigenvalue centrality metric used to evaluate the probability that information or knowledge will likely visit a node through a random walk. PageRank is calculated through an iterative algorithm in which nodes collect PageRank values from their neighbors in every iteration. For simplicity, each node is initially assigned a value of ${\rm PR}(i)=1$. During each iteration, node $i$ collects a PageRank value through the link pointed from its neighbor $j$ ($j\rightarrow i$) as the PageRank of an adjacent node divided by its outbound degree $k^{j}_{\rm out}$. Namely,
\begin{equation} 
{\rm PR(i)} = (1-d) + \sum_{j\in(\partial i \rightarrow
  i)}\frac{{\rm PR}(j)}{k^{j}_{\rm out}}.
\end{equation}

Here $\partial i \rightarrow i$ is the set of points which have outbound links to $i$, and $d$ is a damping factor which we choose as 0.7 in our work. When a converging threshold ($10^{-4}$) is reached, the iteration stops and outputs the final result of PageRank.

Although PageRank was originally proposed for ranking websites, it has also been applied in social network analysis. Given the assumption that senders of messages or makers of phone calls are likely to be the ones providing the information being communicated, PageRank is a good metric to evaluate the likelihood that an individual captures the information spreading in the network. Similarly to k-shell, PageRank requires the global information of the whole network. However, it is easy to update when the network changes.

4. Collective Influence (CI) is an algorithm to identify the most influential nodes via optimal percolation \cite{CI}. Rather than the above heuristic metrics, Collective Influence is introduced by a theoretical approximation of the solution to a problem of influence maximization in locally tree-like social networks \cite{kempe}. CI minimizes the largest eigenvalue of a modified non-backtracking matrix of the network in order to find the minimal set of nodes to disintegrate the network. It has been shown that this process maximizes the spread of information via a threshold model of spreading and also provides the most important nodes for the integrity of the network (optimal percolation). Each node is associated with a CI value, and those with the top CI values are the most influential nodes in the network. The definition of CI is given by:
\begin{equation} 
{\rm CI}(i)=(k_{i}-1)\sum _{j\in \partial \mathrm{Ball}(i,\ell)}{({k}_{j}-1)},
\end{equation}
where the ${\rm Ball}(i,\ell)$ is defined in the text. We should note that the mobile communications network is a typical small world network (average path length $<\ell> \sim 8.9$), and the radius $\ell$ of the ball is limited by the network diameter.

Of the metrics we investigated so far, CI draws our attention since in practice, it has advantages in resolution, correlation with wealth, and scalability to massively large social networks. On the ``global versus local'' issue, we point out that while CI comes from a global theory of maximization of influence, it represents a local approximation in a sphere of influence of finite radius $\ell$. Thus, it is a convenient way to quantify influence in large social networks due to its scalability. Furthermore, in cases where the whole picture of global connectivity is incomplete, the local connectivity up to a few layers $\ell$ might be enough to define network influence and predict the financial status of an individual. On the other hand, we have shown that global quantities like the k-core are also good for capturing an individual's financial status. Indeed, the global k-core contains nested structures of relatively large degrees, which somehow resemble the concentric spheres of influence of a high-CI node. However, the k-core suffers from resolution problems: wealthy people might be located preferentially in the core of the network, but this core is too large to locate them with accuracy. For instance, there are only  25 k-cores in the whole network (Fig. \ref{fig:features}b) to separate one hundred million people, while CI has a larger resolution spanning eight orders of magnitude. Thus, in practical terms, CI presents advantages both in resolution and in high correlation with wealth.

Also, CI represents a balance between a global maximization of
influence and its local approximation in successive layers, allowing
one to use the CI metric in large-scale datasets composed of hundreds
of millions of individuals. Overall, we emphasize that CI is just a
useful strategy for the reasons shown above, but by no means the only
or best way to express the wealth of individuals. More generally,
supervised machine learning can be applied to the problem of
predicting an individual's credit score based on a number of
features. These methods could include not only CI but also the other
measures discussed, along with many other standard network
metrics. Augmenting these measures for determining feature importance
could allow us to better assess which features are important to
determine the wealth of individuals with higher accuracy than that
shown by CI in the present study. The prediction model will give
standard measures of features' importance in further studies when we
have access to more data. Future work will follow this promising
direction.

\section{{Financial parameters and other factors}}

\label{sm4}
We use the following statistics to identify economic effects: First, we separate the individuals into groups on sampling grids in variable space (1D as segment bins and 2D as grids). In each group (with more than 10 people for statistical significance), we count the fraction of wealthy individuals, defined as those individuals in the top 4-quantile $Q>0.75$ or who have a total credit limit greater than USD \$4,000 (converted).

Besides the credit limit, transaction amount and credit score the bank data also provides the information of the clients' birth years. Age as a variable is independent from the network metrics (Supplementary Table \ref{table:tbl1}) and correlates with the percentile-ranking credit limit ($r = 0.42$). However, we do not know the model used by the bank to assign the credit limit, so the age may be a complex reflection of the mixed effects of both increased income and increased account history. Thus, the correlation between age and credit limit might not be capturing only variation in actual wealth but also the amount of information the bank has about the customer.

To quantitatively evaluate the variance caused by network metrics when combined with other factors, we employed Analysis of Covariance (ANCOVA) \cite{ANCOVA}. ANCOVA is an analysis method which conducts regressions between covariate (CV) and dependent variables (DV) under different groups of categorical independent variables (IV). In this case, regression was made between covariate CI and the dependent variable, the fraction of wealth. As in Fig. \ref{fig:features}d, CI is divided into 100 partitions. Based on the information to which we have access, ANCOVA was applied separately among the following independent variables: gender, age, and residential communities. Gender was naturally divided into two groups. Age was grouped year by year from 18 to 65 in a total of 48 groups. The communities were identified by their registered zip code. To reduce the dimensionality of the problem and directly quantify the effect of geographical location, we first sorted the communities by the fraction of wealthy people inside and divided them into 50 balanced groups. We assigned to every community an `Index of Community Wealth' (ICW), which is the quantile ranking of each group that the community belongs to.

The correlation between IVs and CV are shown in Supplementary Table \ref{table:corr}. The negligible correlation between these variables ensures the basic assumption of independence in ANCOVA. Also, in order to test the robustness of our results, the same method was applied under different thresholds of credit limits to define the wealthy population: Q = 0.75 (the threshold we used), 0.85 and 0.95.

The basic output of ANCOVA is a series of p-values showing the significance level of the regression model between CV and DV in different IV groups, and the analysis of variance (ANOVA) \cite{ANCOVA} evaluating the significance of the IVs' effects. The estimated slopes with 95\% confidence intervals are shown in Supplementary Figure \ref{fig:slopes}. Our results show the following: 

1. All IVs' effects are significant ($p<0.001$); namely, the fraction of wealthy people is different among different groups of gender, age or communities. 

2. Inside most groups of each IV, the variation caused by CI is also significant ($p<0.001$). The only exception is that CI's effect is only significant when the clients are older than 24 years (Supplementary Figure \ref{fig:slopes}b). This result indicates that the effect of network metrics, in most cases, is independent from the other known factors. 

3. The slope of regression varies in different groups. However, all the slopes with significant values are positive. 

4. The results of 1 to 3 above are robust under different thresholds of credit line, so Fig. \ref{fig:features} is also similar under different thresholds. Therefore, we focus our results on a given quantile threshold $Q=0.75$ for the remainder of the study. Although the violation of homogeneity in 3 prevents us from making a direct comparison between variables, these results imply that CI significantly and independently affects the fraction of the wealthy population.

\section{Correlation between network metrics and financial status}
\label{sm6}

To compare the value of the social metrics to the economic status of
individuals, we have to draw out the best one to describe network
location influence effects. We sum up all the age groups and consider
the effect of network metrics to demonstrate the effects of each
variable.

\begin{figure} 
\includegraphics[width=0.7\textwidth]{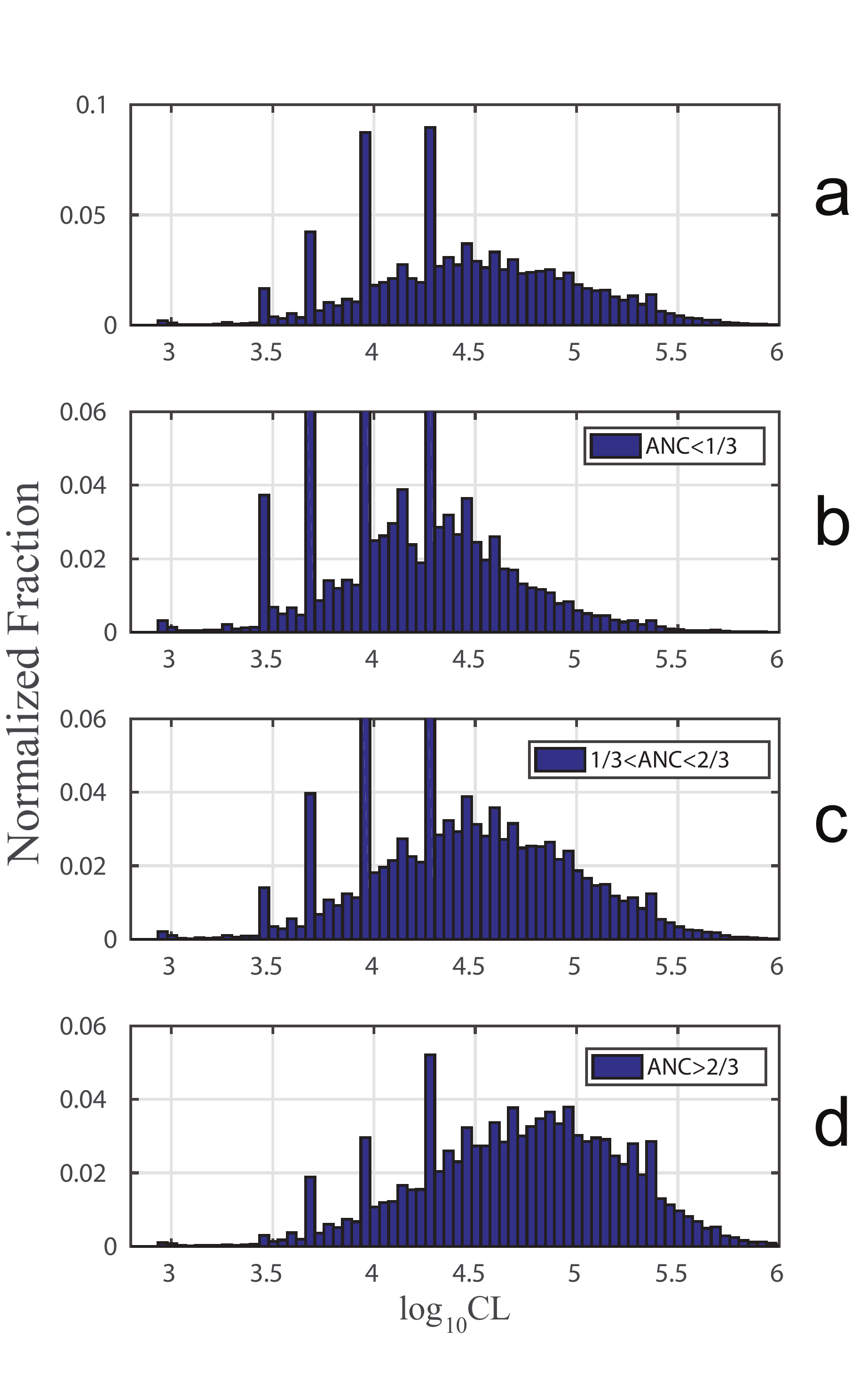}
\caption{{\bf Distribution of Credit Limit (CL) under different age-network composite (ANC) groups.} The distribution is not invariant under changes in ANC. } 

\label{fig:CLDist}
\end{figure}

The reason for using the aggregated model instead of the direct correlations at the individual level is because the regression models at the individual level are based on certain assumptions that are not satisfied by our data. Thus, we were unable to apply regression models at the individual level, and instead provide data at an aggregated level. The failure of regression models at the individual level is due to two reasons: 

1. The distribution of credit limit (CL) for a given level of ANC [which is a log-normal-like distribution with several peaks located at integers such as 50,000 or 100,000 (Supplementary Figure \ref{fig:CLDist}a)] is not invariant under changes in ANC. That is, the distribution changes shape when ANC increases, showing an increasing fraction of high-CL population while the fraction of people around the mean value stays unchanged (Supplementary Figures \ref{fig:CLDist}b--d). Such behavior directly violates the constant variance assumption of regression models and causes the data to be poorly captured by least-square regression models.

2. Besides the above fluctuations in the credit limit, other unknown factors may provide random fluctuations in inferring individuals' financial status. Such combined random effects are considerable at the individual level. However, aggregation models reduce the fluctuation caused by random factors, and the effect of the network emerges at the population level.

Thus, we adjust our statistical model to reflect the complexity of economic effects from network metrics and aggregate the data as follows:

First we separate the individuals into groups of sampling grids in a variable space (in 1D as segment bins and in 2D as grids). In each group (with more than 10 people for statistical significance), we count the fraction of wealthy individuals defined as those individuals in the top 4-quantile $Q>0.75$ or who have a total credit limit greater than (equivalent to) USD \$4,000. The dependence of our results on different wealth thresholds is provided in Supplementary Note \ref{sm4}.

Besides the degree, the volume of communication may have correlations with economic status since we could not eliminate the systematic bias caused by phone call service fees. We investigate the correlation between the fraction of wealthy people and the average communication load per link: ${\rm AVL}_i = \frac{W_i}{k_i}$, where $W_i$ is the volume of communication events and $k_i$ is the degree of node $i$. The regression result shown in Supplementary Figure \ref{fig:AVL} shows that there is no significant correlation between the average communication volume per link and the fraction of wealthy individuals. Therefore, the effect of communication volume is negligible in comparison with the other variables considered in this study.

Supplementary Figure \ref{fig:oneVarReg} shows the results. The large fluctuation in degree for higher quantiles in Supplementary Figure \ref{fig:oneVarReg}a implies that the effect of degree involves complex social patterns rather than only the local properties of the degree of the node. Thus, we abandon the use of degree for further study as an indicator. k-shell is good enough to present a positive correlation of high network location influence. However, due to the limited values of k-core, it cannot provide finer resolution for prediction (Supplementary Figure \ref{fig:oneVarReg}b). Therefore, k-shell is also not considered for further studies as an indicator. The performance of PageRank (Supplementary Figure \ref{fig:oneVarReg}c) with a slightly negative correlation suggests that it is not the optimal variable to rank economic status, and thus it is not considered herein.

\begin{figure} 
\includegraphics[width=\textwidth]{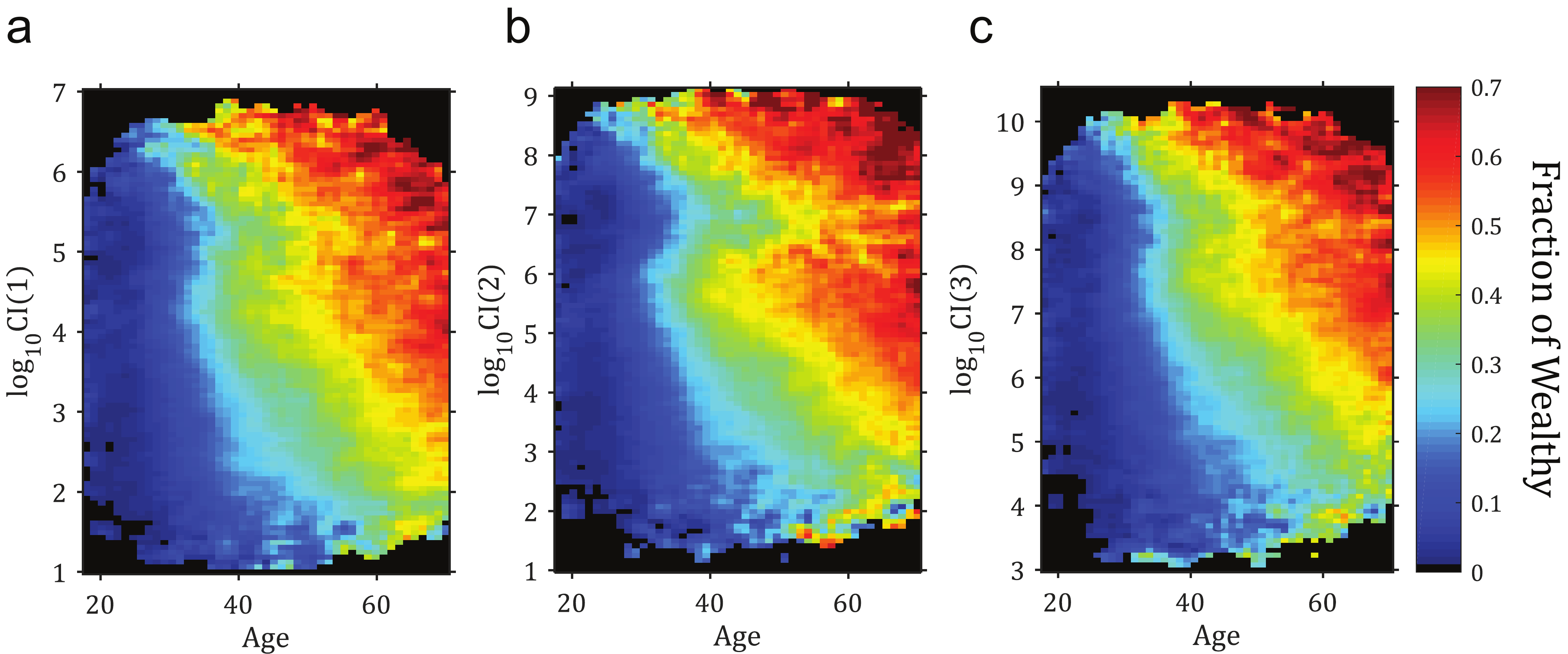}
\caption{{\bf Fraction of wealthy people in each group against age and logarithm collective influence for different radius.} Radii $\ell$ range from 1 to 3. Communities are determined by 200 segments covering from the bottom 1\% to top 1\% of CI values. Only those groups with population larger than 10 are shown on the plot.}

\label{fig:CI123}
\end{figure}

Finally, CI (Supplementary Figure \ref{fig:oneVarReg}d) shows strong global correlation and satisfying resolution, which makes it a convenient metric for quantifying the influence of network location. The strong correlation with CI is invariant under different radii of influence $\ell$ (Supplementary Figure \ref{fig:CI123}).

We notice a non-monotonic oscillatory behavior of the fraction of wealthy people when using $k$ and CI as variates (Supplementary Figures \ref{fig:oneVarReg}a and \ref{fig:oneVarReg}d). This effect is complex and cannot be captured by either the degree or CI, and may not be limited to local properties. The oscillation is reduced when using CI in the analysis, and this is one of our reasons for choosing CI as a potential predictor. We will continue investigating the non-monotonic pattern in future work.

\section{Modularity and Diversity ratio}
\label{sm7}

\begin{figure} 
\includegraphics[width=\textwidth]{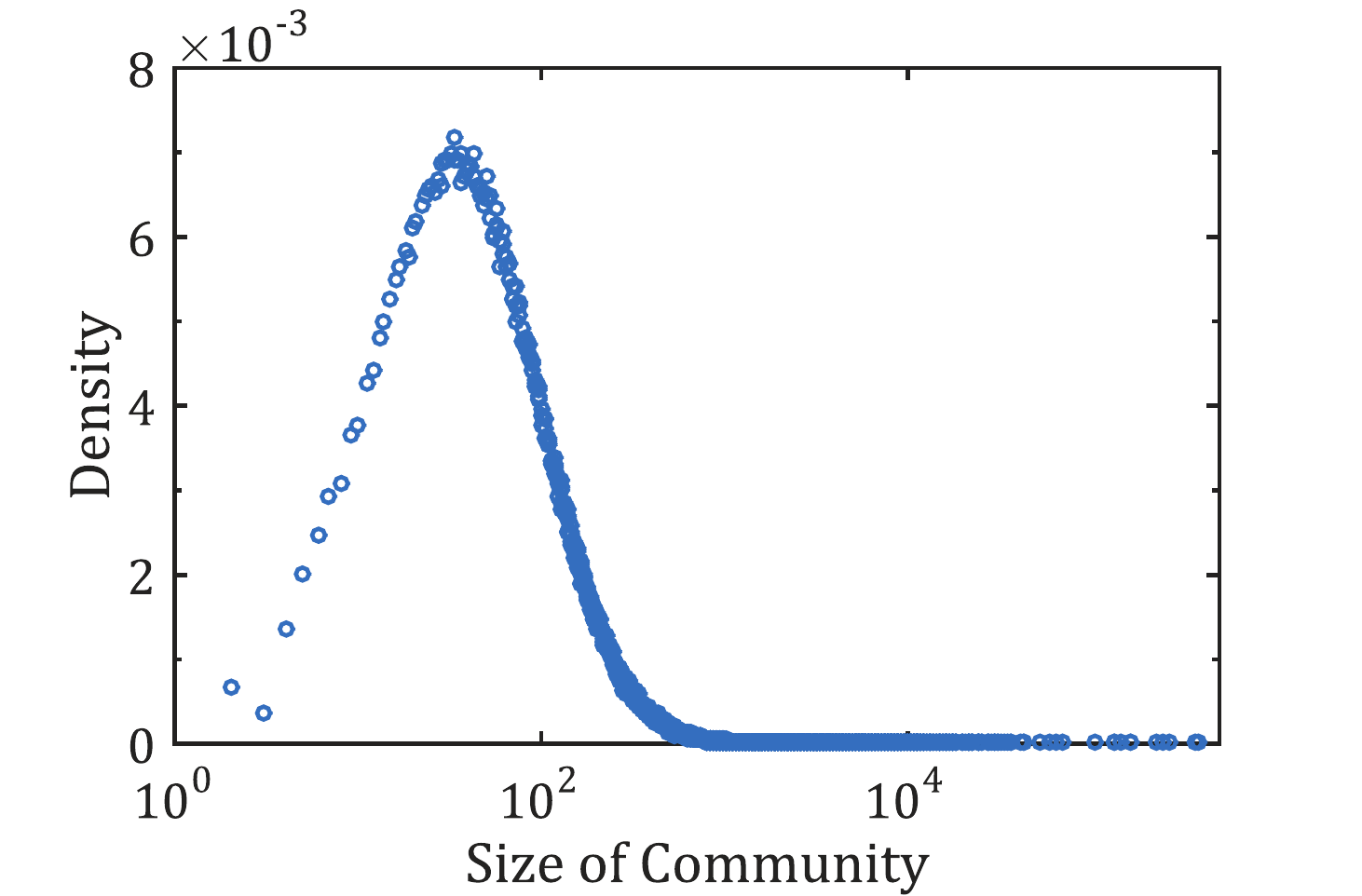}
\caption{{\bf Distribution of community sizes in the entire social network at second iteration.}}

\label{fig:distComW2}
\end{figure}

Additional research on modularity was implemented as follows. Personal structural hole \cite{burt2009structural} effects were evaluated by the ratio of total weights attached with nodes outside a community $k_{\rm out}$, to those inside a community $k_{\rm in}$. A fast community detection algorithm introduced by Blondel {\it et al.} \cite{comDtct} was implemented in this work. The algorithm aims to maximize the modularity function \cite{comDtct,comDtct12}:
  \begin{equation}
    Q_m = \frac{1}{W}\sum_{i,j}[W_{ij}-\frac{W_iW_j}{W}]\delta(c_i,c_j),
  \end{equation}
where $W_{ij}$ is the number of communication events loaded on link ${i,j}$ and $c_i$ is the community label of node $i$. $W_i = \sum_{j\in\partial i} W_{i,j}$ and $W = W_{ij}\sum_{i,j}$. The global maximization of modularity was achieved by iteratively calculating the local maximization of normalized networks based on communities. Different communities were labeled during each iteration. Among all the communities, we chose the clustering of the second iteration to control the average scale of the community to ~$10^2$. There are $4.92\times10^5$ communities inside the network. The distribution of community sizes is fat-tailed with a largest community size of ~$10^6$ (Supplementary Figure \ref{fig:distComW2}). The fraction of wealthy individuals inside each community is independent of the size of the community ($r<0.05$).

After we label the network with its communities, we can evaluate an individual's structural hole effect \cite{burt2009structural} by introducing the diversity ratio DR. DR is defined by the ratio of total communication events with people outside one's own community $W_{\rm out}$ to those with people inside the community, namely $W_{\rm in}$, DR =$ W_{\rm out}/W_{\rm in}$. The ratio is weakly correlated with CI ($r = 0.4$). The same statistic of composite ranking was implemented as CI with the same number of statistic segments and composite factor $\alpha=0.5$ as in the text. The result (Fig. \ref{fig:ageCellCIinout}d) shows that the structural hole effect also has a strong correlation with the distribution of affluent individuals while it is weakly dependent on CI. This result confirms the importance of the ability to communicate with outside communities via ``weak ties'' for personal economic development \cite{granovetter2005impact}.

\section{Marketing Campaign}{}

\label{smmarket}
In the marketing campaign, clients were approached by SMS messages offering a benefit. In the text we sent during the campaign, we did not provide a specific product. Instead, the only information we provided was to notify the client that he/she was eligible for an offer from the bank. This somehow eliminated the bias caused by the nature of a product which may have a different appeal to wealthy or poor people. We sent the following messages:

\begin{blockquote}
Request your credit card with benefits from (Bank\_name) by calling at (Bank\_phone\_number). Fees and requirements at (Bank\_url).

\noindent (Bank\_name) has a special offer for you. If you're interested call at (Bank\_phone\_number). Fees and requirements at (Bank\_url).

\noindent (Bank\_name) has a credit card fit for you. Request it by calling at (Bank\_phone\_number). Fees and requirements at (Bank\_url).

\noindent (Bank\_name) has a credit card with benefits. Request it at (Bank\_phone\_number). Fees and requirements at (Bank\_url).

\noindent (Bank\_name) offers you a credit card with benefits. Request it by calling at (Bank\_phone\_number). Fees and requirements at (Bank\_url).

\noindent (Bank\_name) has an exclusive offer for you, call at (Bank\_phone\_number). Fees and requirements at (Bank\_url). 
\end{blockquote}

\newpage

\begin{table}
\centering

\caption{Results of the group entropy analysis for the wealthy population (with quantile ranking $Q>0.95$) and poor ($0.05<Q<0.1$) population.}

\begin{tabular}{|ll|c|c|c|}
\hline
\textbf{}                                         &         & S              & $R_\mathrm{c}(\ell)$    & $R_\mathrm{n}(\ell)$            \\ \hline
\multicolumn{1}{|l|}{\multirow{2}{*}{$\ell = 1$}} & wealthy & 6.37$\pm$0.12  & 5.5$\pm$0.4    & 9.3$\pm$0.7            \\ \cline{2-5} 
\multicolumn{1}{|l|}{}                            & poor    & 6.68$\pm$0.10  & 4.3$\pm$0.3    & 7.1$\pm$0.5            \\ \hline
\multicolumn{1}{|l|}{\multirow{2}{*}{$\ell = 2$}} & wealthy & 7.94$\pm$0.10  & 141.3$\pm$4.7  & $6.3\pm0.2\times10^2$  \\ \cline{2-5} 
\multicolumn{1}{|l|}{}                            & poor    & 8.38$\pm$0.14  & 101.6$\pm$3.4  & $3.1\pm0.1\times10^2$  \\ \hline
\multicolumn{1}{|l|}{\multirow{2}{*}{$\ell = 3$}} & wealthy & 9.11$\pm$0.11  & 443.0$\pm$11.5 & $7.6\pm0.4\times10^3$  \\ \cline{2-5} 
\multicolumn{1}{|l|}{}                            & poor    & 9.30$\pm$0.12  & 390.9$\pm$6.0  & $4.9\pm0.4\times10^3$  \\ \hline
\multicolumn{1}{|l|}{\multirow{2}{*}{$\ell=4$}}   & wealthy & 10.23$\pm$0.02 & 565.4$\pm$10.7 & $5.10\pm0.04\times10^4$\\ \cline{2-5} 
\multicolumn{1}{|l|}{}                            & poor    & 10.23$\pm$0.04 & 517.0$\pm$9.0  & $4.23\pm0.05\times10^4$\\ \hline
\end{tabular}

\label{table:entropy}
\end{table}

\begin{table}
\centering
\medskip
\caption{Correlation ($r$-values) between the metric centralities obtained from the social network and age.}
\begin{tabular}{|c | c|c|c|c|}
	\hline & $k$ & k-shell & PageRank & $\log_{10}$CI \\ \hline
        Age & -0.021 & -0.016 & -0.033 & -0.007 \\\hline $k$ & & 0.972 &
        0.648 & 0.953 \\\hline k-shell & & & 0.589 & 0.960 \\\hline PageRank & &
        & & 0.575 \\ 
\hline
\end{tabular}

\label{table:tbl1}
\end{table}

\begin{table}
\centering
\caption{{\bf Correlation between covariate CI and independent variables: age, gender and Index of Community Wealth (ICW).}
The correlation between gender and other features is presented through
the Point-Biserial correlation coefficient, and other correlations are
Pearson correlations. Point-Biserial correlation coefficients quantify
the male as 1 and female as 0 and are defined as: $r =
\frac{\bar{X_1}-\bar{X_0}}{s_{n-1}}\sqrt{\frac{n_1n_0}{n(n-1)}}$. $n$
is the total number of samples. $n_1$ and $n_0$ refer to the
population inside each group. $\bar{X_1}$ and $\bar{X_0}$ are the
means of the variables in each group. $s_{n-1}$ is the estimated
unbiased standard deviation of $X$: $s_{n-1} =
\sqrt{\frac{1}{n-1}\sum^n_{i=1}(X_i-\bar{X})^2}$.
}
\begin{tabular}{|l|l|l|l|}
\hline
        & CI      & Gender  & ICW     \\ \hline
Gender  & -0.0419 &         &         \\ \hline
ICW     & -0.0093 & 0.0131  &         \\ \hline
Age     & -0.0007 & -0.0116 & -0.0022 \\ \hline
\end{tabular}
\label{table:corr}
\end{table}

\begin{figure} 
\includegraphics[width=\textwidth]{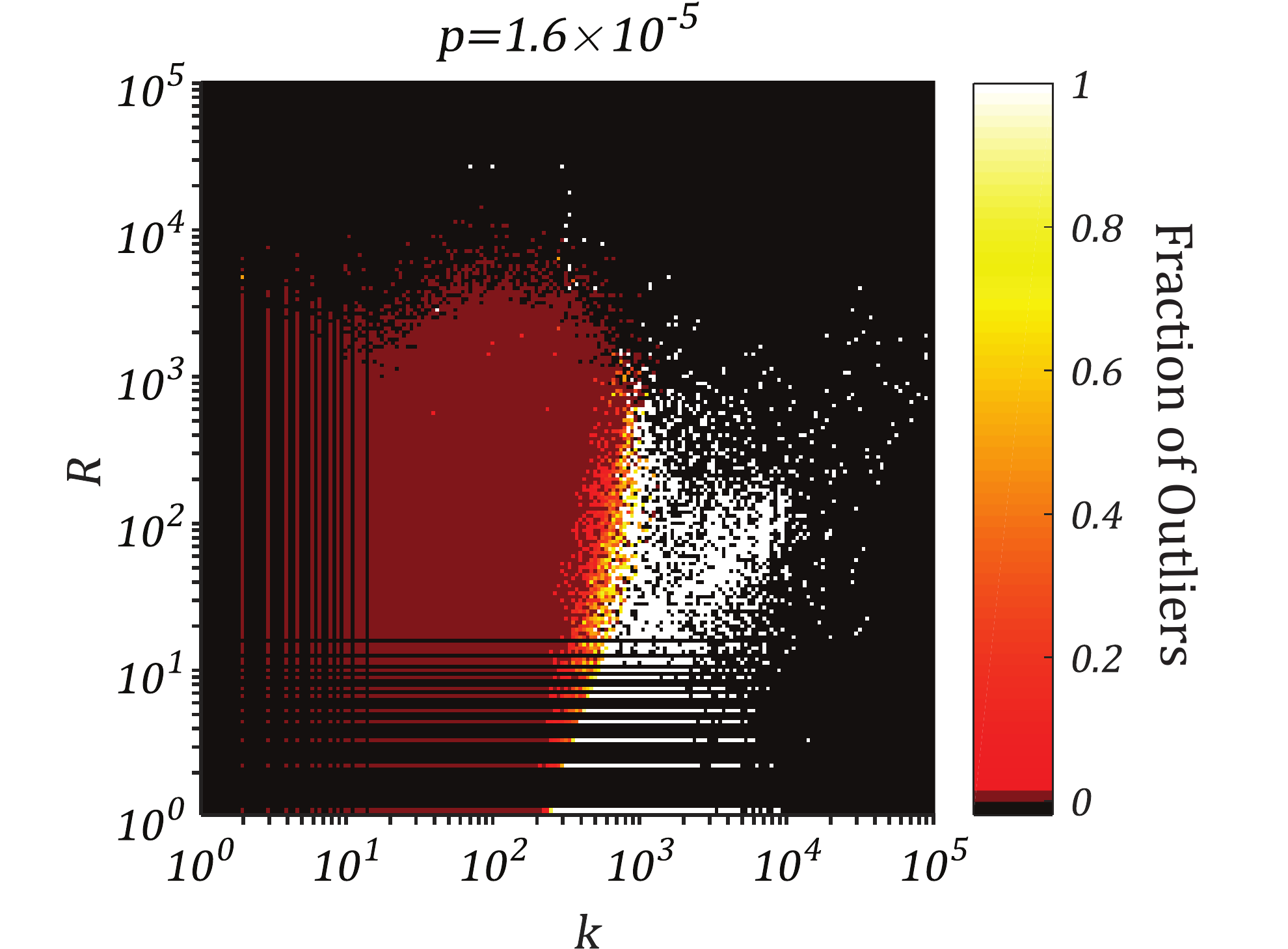}
\caption{{\bf Final result of data filtering.} The result is presented in the space of $k$ and communication pairs $R$. The data points were put into a grid bin of 200$\times$200. The color represents the fraction of outliers in each bin. The filter gives us a gradual boundary of human- and non-human-operated lines. }

\label{fig:filResult}
\end{figure}

\begin{figure} 
\includegraphics[width=\textwidth]{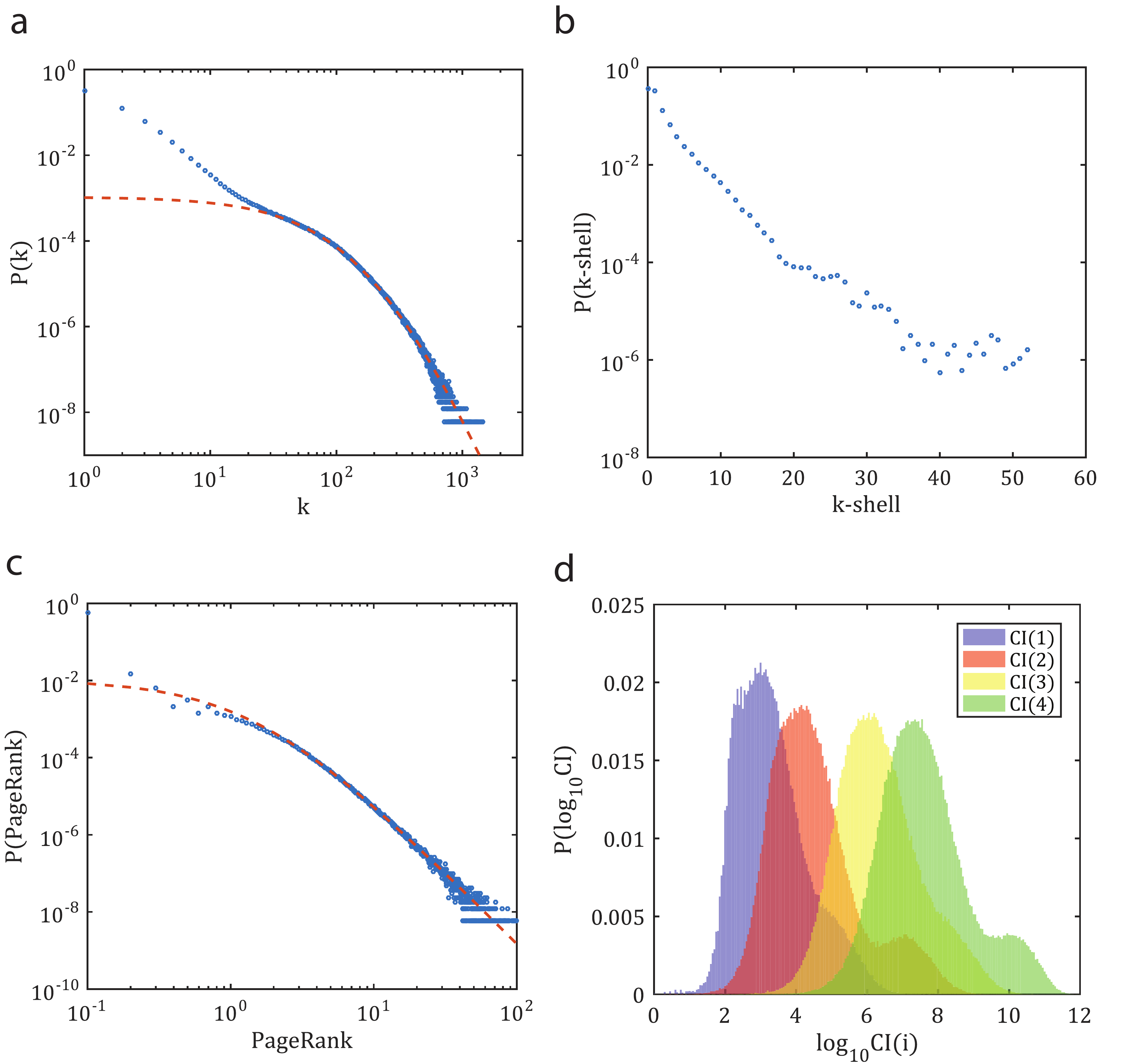}
\caption{{\bf Distribution of network metrics.} \textbf{(a)} degree, \textbf{(b)} k-core, \textbf{(c)} PageRank, and \textbf{(d)} Collective Influence ($\ell=$1 to 4). Collective Influence follows a double-tailed distribution. A small peak for larger CI emerges for even $\ell$.}

\label{fig:Distri}
\end{figure}

\begin{figure} 
\includegraphics[width = 0.7\textwidth]{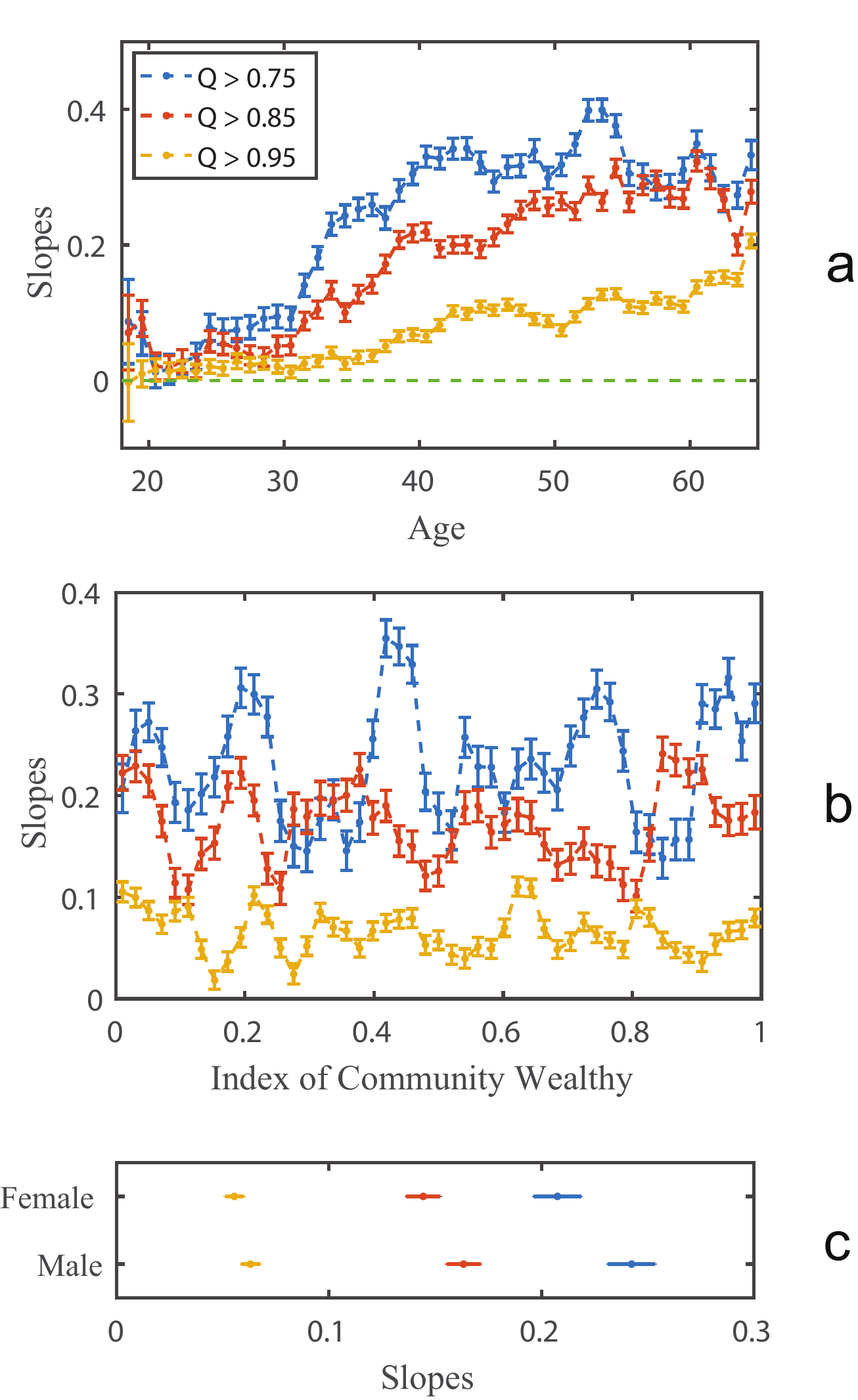}
\caption{{\bf Estimated slopes in different groups of independent variables.} {\textbf (a)}, Age, {\textbf (b)}, Index of Community Wealth (ICW), and {\textbf (c)}, Gender. $95\%$ confidence interval is marked by error bars in the plot. Different thresholds of wealth $Q$ are labeled by different colors.}

\label{fig:slopes}
\end{figure}

 \begin{figure}
\includegraphics[width=\textwidth]{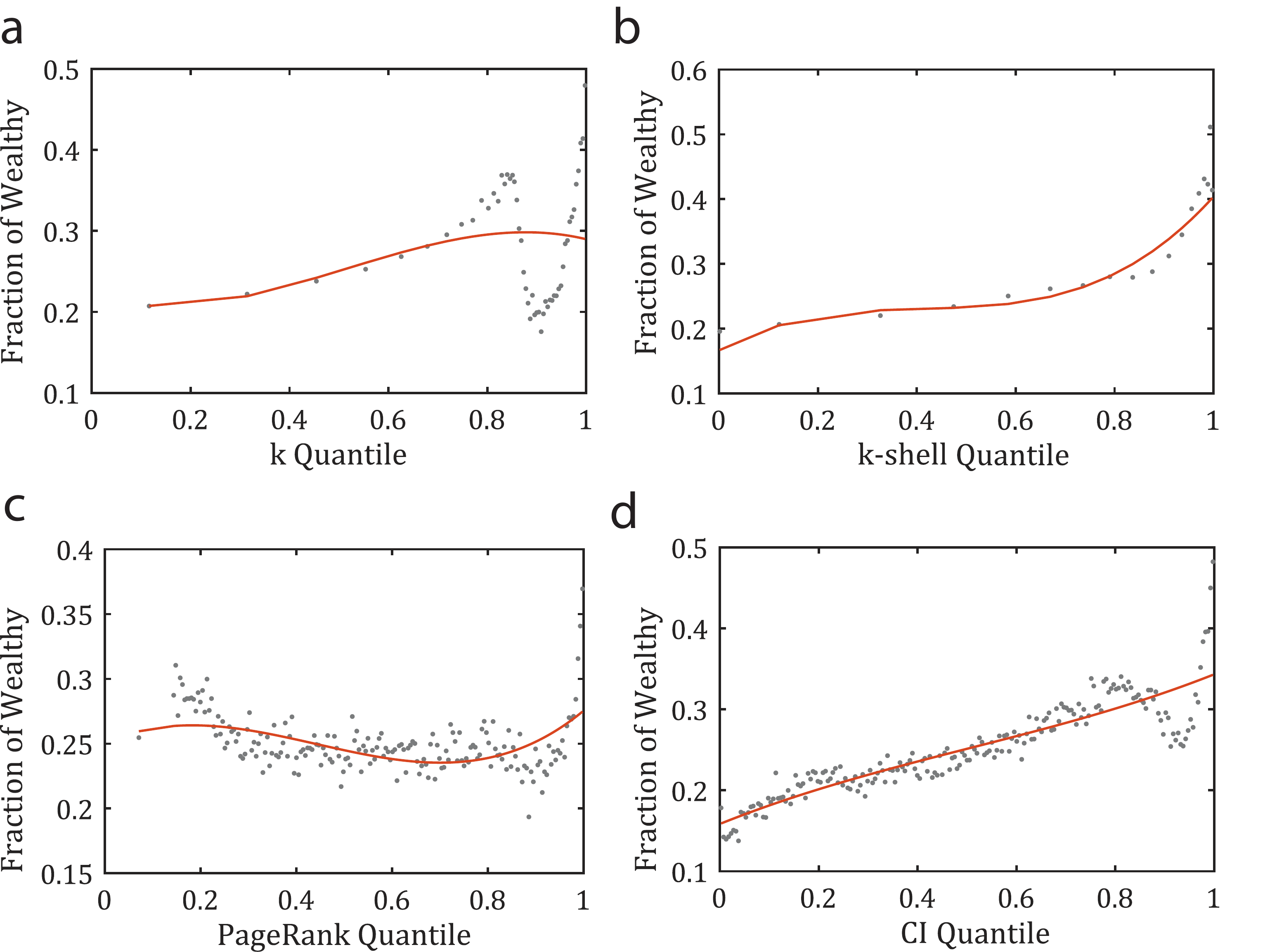}
\caption{{\bf Fitting results of wealthy population vs. network influence metrics along with corresponding $R^2$ values.} \textbf{(a)} Degree (0.51), \textbf{(b)} k-core (0.99), \textbf{(c)} PageRank (0.28), and \textbf{(d)} Collective Influence (0.80). All variables are normalized to [0, 1] by the quantile ranking to ensure an adequate number of data points in each partition. The entire quantile ranking is divided into 200 segments from minimum to maximum. Only those groups with population larger than 10 are shown on the plot. Out of the four metrics, CI is the most convenient for capturing high correlations and presenting a large range of values that allow us to classify the whole population. }

\label{fig:oneVarReg}
\end{figure}

\begin{figure} 
\includegraphics[width=0.7\textwidth]{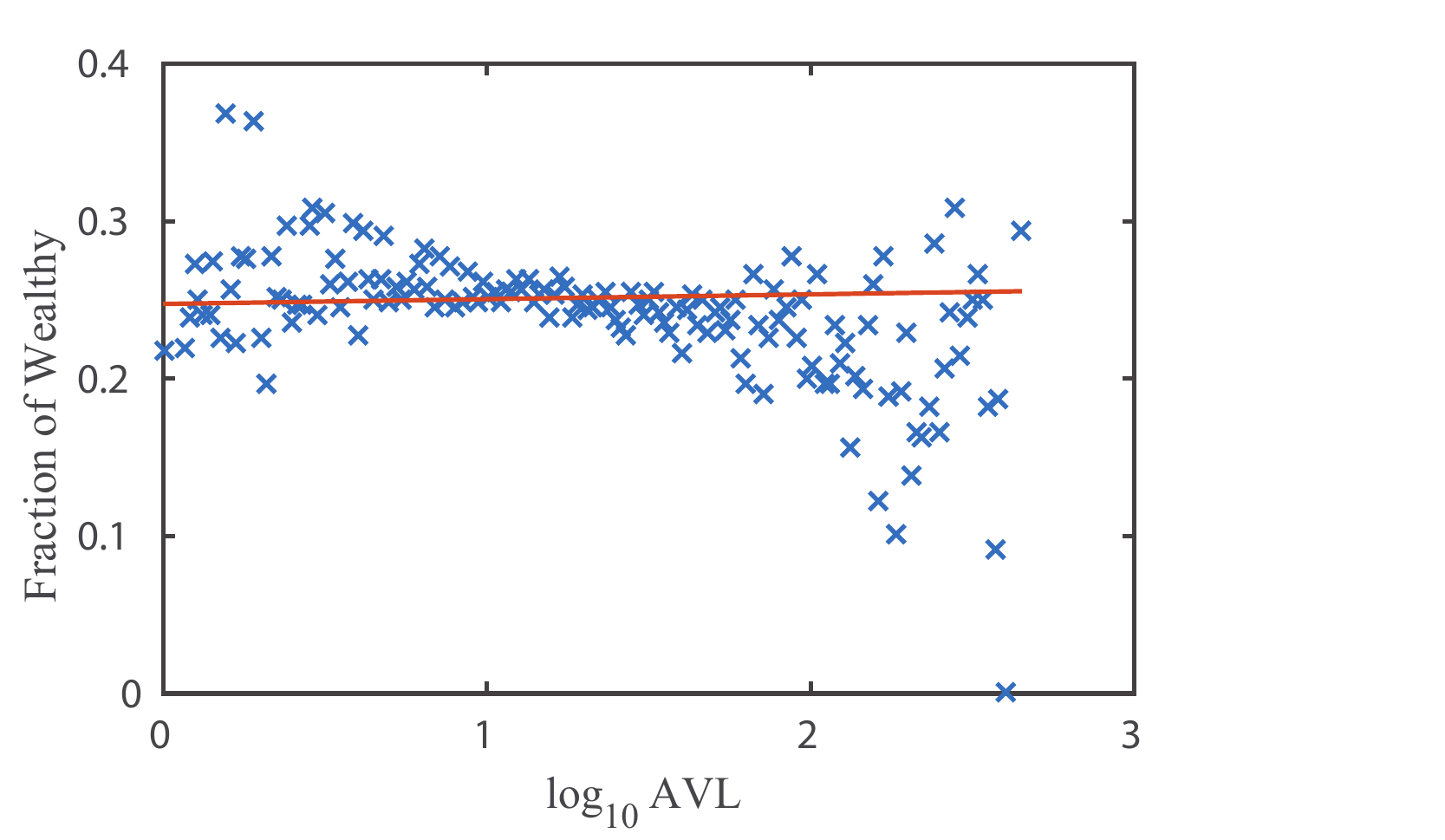}
\caption{{\bf Fraction of wealthy people vs. average communication event load per link (AVL).} AVL is in log-10 scale and divided into 200 partitions. Each group with a population of more than 10 is considered in counting the fraction of wealthy people inside the group.}

\label{fig:AVL}
\end{figure}

\clearpage

{\bf Supplementary References}

\end{document}